\documentclass[11pt,epsf]{article}
 \usepackage{amsmath}
 \usepackage{graphicx}
 \usepackage[merge,numbers,compress]{natbib}
 \usepackage[T1]{fontenc}
 \usepackage{booktabs}
 \usepackage{xcolor} 
 \usepackage{xspace}
 \usepackage{dcolumn}
 \usepackage{placeins}
 \usepackage{amssymb}
 \usepackage[colorlinks=true,citecolor=blue!50!black,linkcolor=black]{hyperref}
 \usepackage{caption}
 \usepackage
 [subrefformat=parens,position=top,skip=-15pt,margin=15pt,justification=justified,singlelinecheck=false]
 {subcaption}
 \usepackage{array,multirow}

\newcommand{\Pl}{\ell}

\newcommand{\pb}{{\ensuremath\unskip\,\text{pb}}\xspace}

\def\be{\begin{equation}}
\def\ee{\end{equation}}

\newcommand{\Pj}{\ensuremath{\text{j}}\xspace}
\newcommand{\Pp}{\ensuremath{\text{p}}\xspace}

\newcommand{\Pb}{\ensuremath{\text{b}}\xspace}

\newcommand{\Pu}{\ensuremath{\text{u}}\xspace}
\newcommand{\Pd}{\ensuremath{\text{d}}\xspace}
\newcommand{\Ps}{\ensuremath{\text{s}}\xspace}
\newcommand{\Pc}{\ensuremath{\text{c}}\xspace}
\newcommand{\Pg}{\ensuremath{\text{g}}\xspace}

\newcommand{\PW}{\ensuremath{\text{W}}\xspace}
\newcommand{\PZ}{\ensuremath{\text{Z}}\xspace}

\newcommand{\MWOS}{\ensuremath{M_\PW^\text{OS}}\xspace}
\newcommand{\MW}{\ensuremath{M_\PW}\xspace}
\newcommand{\MZOS}{\ensuremath{M_\PZ^\text{OS}}\xspace}
\newcommand{\MZ}{\ensuremath{M_\PZ}\xspace}

\newcommand{\GZOS}{\ensuremath{\Gamma_\PZ^\text{OS}}\xspace}

\newcommand{\GWOS}{\ensuremath{\Gamma_\PW^\text{OS}}\xspace}

\newcommand{\GeV}{\ensuremath{\,\text{GeV}}\xspace}
\newcommand{\TeV}{\ensuremath{\,\text{TeV}}\xspace}

\newcommand{\alphas}{\ensuremath{\alpha_\text{s}}\xspace}

\newcommand{\GF}{\ensuremath{G_\mu}}

\newcommand{\ptsub}[1]{\ensuremath{p_{\text{T},#1}}\xspace}

\newcommand{\MVOS}{\ensuremath{M_{\text{V}}^\text{OS}}\xspace}%
\newcommand{\GVOS}{\ensuremath{\Gamma_{\text{V}}^\text{OS}}\xspace}%

\newcommand{\newc}{\newcommand}
\newc{\bi}{\begin{itemize}}
\newc{\ei}{\end{itemize}}
\newc{\benu}{\begin{enumerate}}
\newc{\eenu}{\end{enumerate}}
\newc{\bc}{\begin{center}}
\newc{\ec}{\end{center}}
\newc{\bfig}{\begin{figure}}
\newc{\efig}{\end{figure}}
\newc{\qbar}{\bar{q}}
\newc{\go}{\tilde{g}}
\newc{\PB}{\textsc{Powheg-Box}}

\newcommand{\madgraph}{{\sc\small MadGraph5\_aMC@NLO}\xspace}

\newcommand{\rT}{{\mathrm{T}}}
\newcolumntype{.}{D{.}{.}{-1}}
\newcolumntype{d}[1]{D{.}{.}{#1}}

\newcommand{\QCD}{\ensuremath{\text{QCD}}}

\newcommand{\mr}[1]{\ensuremath{\mathrm{#1}}}

\newcommand{\muR}{\ensuremath{\mu_{\mr{R}}}}
\newcommand{\muF}{\ensuremath{\mu_{\mr{F}}}}

\colorlet{tableoverheadcolor}{gray!37.5}
\colorlet{tableheadcolor}{gray!25}
\colorlet{tablerowcolor}{gray!12.5}

\newlength{\width}
\newlength{\height}

\def\draftdate{\relax}
\def\mda{\relax}
\def\mua{\relax}
\def\mla{\relax}
\def\draft{
\def\thtystars{******************************}
\def\sixtystars{\thtystars\thtystars}
\typeout{}
\typeout{\sixtystars**}
\typeout{* Draft mode!
         For final version remove \protect\draft\space in source file *}
\typeout{\sixtystars**}
\typeout{}
\def\draftdate{\today}
\def\mua{\marginpar[\boldmath\hfil$\uparrow$]%
                   {\boldmath$\uparrow$\hfil}\color{black}%
                    \typeout{marginpar: $\uparrow$}\ignorespaces}
\def\mda{\color{red}\marginpar[\boldmath\hfil$\downarrow$]%
                   {\boldmath$\downarrow$\hfil}%
                    \typeout{marginpar: $\downarrow$}\ignorespaces}
\def\mla{\marginpar[\boldmath\hfil$\rightarrow$]%
                   {\boldmath$\leftarrow $\hfil}%
                    \typeout{marginpar: $\leftrightarrow$}\ignorespaces}
\def\Mua{\marginpar[\boldmath\hfil$\Uparrow$]%
                   {\boldmath$\Uparrow$\hfil}\color{black}%
                    \typeout{marginpar: $\uparrow$}\ignorespaces}
\def\Mda{\color{red}\marginpar[\boldmath\hfil$\Downarrow$]%
                   {\boldmath$\Downarrow$\hfil}%
                    \typeout{marginpar: $\downarrow$}\ignorespaces}
\def\Mla{\marginpar[\boldmath\hfil\textcolor{red}{$\Rightarrow$}]%
                   {\boldmath\textcolor{red}{$\Leftarrow $}\hfil}%
                    \typeout{marginpar: $\leftrightarrow$}\ignorespaces}
\overfullrule 5pt
\oddsidemargin 15mm
\marginparwidth 29mm
}

\begin{document}

\title{\hfill ~\\[-30mm]
\phantom{h} \hfill\mbox{\small CAVENDISH--HEP--22/09, FR-PHENO-2022-09, P3H-22-117, TTK-22-41}
\\[1cm]
\vspace{13mm}   \textbf{A detailed investigation of W+c-jet at the LHC}}

\date{}
\author{
Micha\l{} Czakon$^{1\,}$\footnote{E-mail:  \texttt{mczakon@physik.rwth-aachen.de}},
Alexander Mitov$^{2\,}$\footnote{E-mail:  \texttt{adm74@cam.ac.uk}},
Mathieu Pellen$^{3\,}$\footnote{E-mail:  \texttt{mathieu.pellen@physik.uni-freiburg.de}},
Rene Poncelet$^{2\,}$\footnote{E-mail:  \texttt{poncelet@hep.phy.cam.ac.uk}}
\\[9mm]
{\small\it $^1$ Institut f{\"u}r Theoretische Teilchenphysik und Kosmologie, RWTH Aachen University,} \\ %
{\small\it RWTH Aachen University, D-52056 Aachen, Germany}\\[3mm]
{\small\it $^2$ Cavendish Laboratory, University of Cambridge,} \\ %
{\small\it J.J. Thomson Avenue, Cambridge CB3 0HE, United Kingdom}\\[3mm]
{\small\it $^3$ Albert-Ludwigs-Universit\"at Freiburg, Physikalisches Institut,} \\ %
{\small\it Hermann-Herder-Stra\ss e 3, D-79104 Freiburg, Germany}\\[3mm]
        }
\maketitle

\begin{abstract}
\noindent

State-of-the-art analyses of W+c-jet production at the LHC require precise predictions. 
In the present work, we study in detail the impact of off-diagonal CKM elements up to next-to-next-to leading order in QCD, the influence of flavored jet algorithms, and the size of electroweak corrections.
In addition, we also investigate phenomenological aspects related to the exact definition of the process.
We find that all these effects can be of the order of several per cent for both the fiducial cross section and differential distributions.
They are, therefore, very relevant for the interpretation of current and upcoming measurements.

\end{abstract}
\thispagestyle{empty}
\vfill

\newpage

\section{Introduction}

The extraordinary precision of the Large Hadron Collider (LHC) allows to investigate in detail the fundamental structure of elementary particles.
A prime example is the strange-quark content of the proton whose asymmetry has been predicted at three loop in QCD~\cite{Catani:2004nc}.
In the past, the strange-quark parton distribution function (PDF) has been determined by non-LHC experiments~\cite{Lai:2007dq,Faura:2020oom}.
Nowadays, it can be constrained from the measurement of W+c-jet at the LHC~\cite{Baur:1993zd} and several such measurements have already been performed~\cite{Aad:2014xca,CMS:2018muk,Sirunyan:2018hde,CMS:2019rlx,CMS:2021oxn,CMS:2022bjk}.

The basic idea is that at the Born level, the strange-quark PDF is directly related to the cross section of the process.
A charm quark in the final state, implies a strange quark in the initial state (see left of Fig.~\ref{fig:diag}).
The inclusion of non-diagonal CKM elements (see middle and right of Fig.~\ref{fig:diag}) or higher-order QCD corrections (see Table 1 of Ref.~\cite{Czakon:2020coa}), however, renders this picture significantly more complex.
In order to benefit from new experimental measurements, precise theory predictions are required for their interpretations.

The next-to-leading order (NLO) QCD cross sections for W+c-jet production at the Tevatron \cite{Giele:1995kr} and at the LHC \cite{Stirling:2012vh} have been known for a long time.
More recently, Ref.~\cite{Bevilacqua:2021ovq} went beyond this by computing NLO QCD corrections matched to parton shower with massive charm quarks.
In Ref.~\cite{Czakon:2020coa}, a first computation of next-to-next-to leading order (NNLO) QCD corrections has been presented.
In that reference, off-diagonal CKM elements were included only at leading order (LO) and the flavored $k_\rT$ algorithm was used.
Finally, while electroweak (EW) corrections are known for inclusive W+j production~\cite{Kuhn:2007qc,Kuhn:2007cv,Hollik:2007sq,Denner:2009gj}, they were still unknown for W+c production and thus were not included in the predictions of Ref.~\cite{Czakon:2020coa}.

In the present work, we extend the computation of Ref.~\cite{Czakon:2020coa} by presenting the first NNLO QCD calculation of W+c production at the LHC with full CKM dependence.
We also compute the dominant NLO EW corrections for this process for the first time.
In addition, we study in detail the numerical effect of the charm-jet definition.
This is particularly important since so far infrared(IR)-safe computations of processes involving flavored jets beyond NLO QCD~\cite{Behring:2019iiv,Gauld:2019yng,Czakon:2020qbd,Gauld:2020deh} have been computed with the flavored $k_\rT$ algorithm \cite{Banfi:2006hf}, while experimental analyses have been carried out with the anti-$k_\rT$ algorithm~\cite{Cacciari:2008gp}.
A fair comparison between theory and experiment, therefore, requires either the use of unfolding corrections or of comparable jet algorithms in both theory and experiment.
We follow the second approach and apply the recent IR-safe flavored anti-$k_\rT$ jet algorithm proposed by some of us~\cite{Czakon:2022wam}.\footnote{We note that alternative proposals have been recently been made in the literature~\cite{Caletti:2022glq,Caletti:2022hnc,Gauld:2022lem}.}
Finally, in addition to these theoretical considerations, we also investigate more phenomenological aspects that are crucial for a theory/experiment comparison.
In particular, we compare several process definitions regarding the charge and multiplicity of charm jets.

The article is organized as follow: in Section~\ref{sec:comp_details}, the details of the calculations are presented.
These include the numerical inputs and the phase-space definition used throughout.
Section~\ref{sec:th} provides our best predictions which include full off-diagonal CKM dependence up to NNLO QCD accuracy and NLO EW corrections.
Section~\ref{sec:details} represents a detailed study of various theoretical aspects such as the flavored jet and process definitions, the significance of off-diagonal CKM elements, scale and PDF dependence. 
Finally, Section~\ref{sec:concl} contains a summary of our main findings and concluding remarks.

\section{Details of the calculations}\label{sec:comp_details}

\subsection{Definition of the process}

\begin{figure}
\centering
\includegraphics[width=0.32\textwidth]{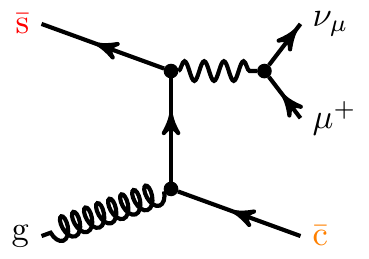}
\includegraphics[width=0.32\textwidth]{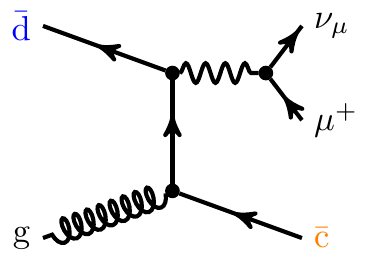}
\includegraphics[width=0.32\textwidth]{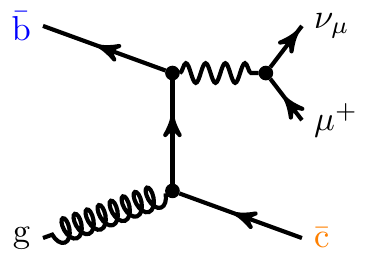}
\caption{Tree-level Feynman diagrams with diagonal CKM elements (left) and off-diagonal elements (centre and right) for $\Pp\Pp \to \mu^+  \nu_{\mu} \Pj_{\rm c}$.}
\label{fig:diag}
\end{figure}

The two processes under investigation are the production of a charm jet in association with an off-shell W boson that decays into a muon and an anti-neutrino (an anti-muon and a neutrino) at the LHC.
The hadronic definition is thus:
\begin{eqnarray}
&& \Pp\Pp \to \mu^+  \nu_{\mu} \Pj_{\rm c} + X\,,\label{eq:w+j}\\
&& \Pp\Pp \to \mu^-  \bar\nu_{\mu} \Pj_{\rm c} + X .\label{eq:w-j}
\end{eqnarray}
At LO, the processes are defined at order $\mathcal{O} \left(\alphas \alpha^2 \right)$ in the strong and EW couplings.
We would like to emphasize that, unless stated otherwise, full dependence on the CKM matrix is kept in all calculations.
In Fig.~\ref{fig:diag}, the three LO diagrams contributing to $\Pp\Pp \to \mu^+  \nu_{\mu} \Pj_{\rm c}$ are presented.
The left diagram is proportional to a diagonal CKM element ($V_{\Pc\Ps}$) while the other two (centre and right) are proportional to off-diagonal elements ($V_{\Pc\Pd}$ and $V_{\Pc\Pb}$, respectively).
Note that in the following, we sometimes refer to the hadronic processes of Eqs.~\eqref{eq:w+j} and \eqref{eq:w-j} as $\Pp\Pp \to \PW^+\Pj_{\rm c}$ and $\Pp\Pp \to \PW^-\Pj_{\rm c}$, respectively.
Nonetheless, off-shell W production is assumed throughout the article.

\begin{paragraph}{QCD corrections}
 At NLO, the QCD corrections include all virtual and real contributions of order $\mathcal{O} \left(\alphas^2 \alpha^2 \right)$. In the same way, at NNLO accuracy all double-virtual, double-real, and real-virtual contributions of order $\mathcal{O} \left(\alphas^3 \alpha^2 \right)$ are taken into account.
 The calculation is carried out in the 5-flavor scheme with massless bottom and charm quarks.
\end{paragraph}

\begin{paragraph}{EW corrections}
 In the present computation we provide NLO EW corrections of order $\mathcal{O} \left(\alphas \alpha^3 \right)$.
 The EW virtual corrections are fully included.
 Thanks to the unitarity of the CKM matrix, the CKM dependence completely factorises in one-loop EW amplitudes and CKM-diagonal matrix elements can be used.
 In the present case, the real corrections involve single real photon emission to cancel the corresponding IR divergences appearing in the EW one-loop amplitude.
 No photon-induced contributions, which constitute an IR finite set, are included.
 The resulting EW corrections have the advantage that they do not depend on the jet algorithm (as they contain only one charm parton in the final state) and thus are insensitive to the event selections regarding the multiplicity and type of c-jet (see Sec.~\ref{sec:event-selection}). They can thus be combined with any of the corresponding QCD corrections presented below.
 Note that the EW corrections have been obtained with the NNPDF3.1 set.
 Using the relative EW corrections with other PDF sets might lead to minor differences.
  
 Subleading NLO corrections of order $\mathcal{O} \left(\alpha^4 \right)$ are neglected here since they have been found to be below a per cent at the level of the cross section for $\Pp\Pp\to\PZ \Pj$ \cite{Denner:2019zfp}.
\end{paragraph}

\subsection{Numerical inputs}

The numerical results presented here are for the LHC with a centre-of-mass energy of $\sqrt{s} = 13 \TeV$.
The numerical values of the CKM elements are the ones from Ref.~\cite{ParticleDataGroup:2020ssz}
\begin{align}
V_{\Pu\Pd} = 0.97401&, \quad V_{\Pu\Ps} = 0.22650, \quad V_{\Pu\Pb} = 0.00361 ,  \nonumber \\ 
V_{\Pc\Pd} = 0.22636&, \quad V_{\Pc\Ps} = 0.97320, \quad V_{\Pc\Pb} = 0.04053 . 
\end{align}

The nominal PDF set used in this computation is NNPDF3.1 with $\alphas=0.118$ \cite{Ball:2017nwa}. To evaluate the PDF uncertainty of the NNPDF3.1 sets, instead of reverting to the 100 replicas provided, we have used specialised minimal PDF sets~\cite{Carrazza:2016htc} which contain only 8 replicas.
In addition, we have computed predictions with the NNPDF3.0~\cite{NNPDF:2014otw} and NNPDF4.0~\cite{NNPDF:2021njg} sets, both with $\alphas=0.118$, and with the CT18NNLO set~\cite{Hou:2019efy}.
The present selection of PDF sets does not include all available PDF sets.
In the future, to make reliable statement about the strange-quark content of the proton, other PDF sets such as MSHT20~\cite{Bailey:2020ooq} and ABMP16~\cite{Alekhin:2018pai} should also be considered.
In particular, they constraint the strange-quark PDFs with different data and assume different functional forms for the strange-quark PDFs.

Regarding the EW input parameters, the electromagnetic coupling is taken in the $G_\mu$ scheme \cite{Denner:2000bj} using the Fermi constant
\begin{equation}
  \alpha = \frac{\sqrt{2}}{\pi} G_\mu \MW^2 \left( 1 - \frac{\MW^2}{\MZ^2} \right)  \qquad \text{with}  \qquad   {\GF = 1.16638\times 10^{-5}\GeV^{-2}}.
\end{equation}
The numerical values of the masses and widths read
\begin{alignat}{2}
\label{eqn:ParticleMassesAndWidths}
                \MZOS &=  91.1876\GeV,      & \quad \quad \quad \GZOS &= 2.4952\GeV,  \nonumber \\
                \MWOS &=  80.379\GeV,       & \GWOS &= 2.085\GeV.
\end{alignat}
The pole masses and widths used in the numerical evaluation are translated from the measured on-shell (OS) values for the massive gauge bosons by \cite{Bardin:1988xt}
\begin{equation}
        M_V = \frac{\MVOS}{\sqrt{1+(\GVOS/\MVOS)^2}}\,,\qquad  
\Gamma_V = \frac{\GVOS}{\sqrt{1+(\GVOS/\MVOS)^2}}.
\end{equation}
The intermediate W-boson resonances are treated in the complex-mass scheme~\cite{Denner:1999gp,Denner:2005fg,Denner:2006ic} in all the computations presented here.
The mass of the charged lepton is set to zero.

Finally, as in Ref.~\cite{Czakon:2020coa}, the common central renormalization ($\muR$) and factorization ($\muF$) scale used is
\begin{equation}
\label{eq:scale}
 \mu_0 = \frac12 \left( E_{\text{T},\PW} + \ptsub{\Pj_\text{c}} \right),
\end{equation}
where $E_{\rm T, \PW} = \sqrt{\MW^2 + \left( \vec{p}_{\rm T,\Pl} + \vec{p}_{\rm T, \nu} \right)^2}$.
To estimate missing higher-order QCD corrections, a 7-points scale variation is performed.

\subsection{Flavored jet algorithms and event selections}
\label{sec:cuts}
 
\subsubsection{Jet algorithms}\label{sec:jet-algos}

 In the present study, we utilize two different flavored jet algorithms.
 This allows for their systematic comparison and for quantifying the effect they have on experimental measurements.
 
The first one, the flavored $k_\rT$ algorithm \cite{Banfi:2006hf}, requires the definition of a beam transverse momentum ($k_{\rT B}$ and $k_{\rT \bar{B}}$). While all pseudo-jets have to be included in the definition, one is free to include or not additional non-QCD particles (W bosons/leptons in the present case) to the beam definition. There is further freedom in deciding whether $\Pc\Pc$ or $\bar\Pc\bar\Pc$ pairs are considered flavored or not. This leads to the so called charge-agnostic and charge-dependent algorithms, defined as follows:
\begin{eqnarray}
{\rm charge~agnostic}~:~&& \left( \sum_i |f_i| \right)( {\rm mod}\,2) \neq 0\,,\nonumber\\
{\rm charge~dependent}~:~&& \sum_i f_i \neq 0\,,
\label{eq:charge-jet-def}
\end{eqnarray}
where $f_i$ is the flavor of parton $i$. The definition Eq.~\eqref{eq:charge-jet-def} implies that a jet containing a $\Pc\Pc$ pair will be treated as flavorless by the charge agnostic algorithm but as flavored by the charge dependent one. In Refs.~\cite{Gauld:2019yng,Czakon:2020coa}, such pairs have been taken to be unflavored based on the argument that experimentally, it is very challenging to determine the charge of the jets in addition to its flavor.
To be able to quantify such effects, in the present work we have considered both cases.

We thus arrive at the following jet definitions
 \begin{itemize}
  \item flavored $k_\rT$ algorithm, charge agnostic (dubbed $k_\rT$CA),
  \item flavored $k_\rT$ algorithm, charge dependent (dubbed $k_\rT$CD),
  \item flavored $k_\rT$ algorithm, charge dependent, with beam definition including W  momenta (dubbed $k_\rT$CDB).
 \end{itemize}

An alternative to this jet algorithm is the flavored anti-$k_\rT$ algorithm~\cite{Czakon:2022wam} which has the advantage that it is almost identical to the standard anti-$k_\rT$ one \cite{Cacciari:2008gp} typically used in experiments. It only requires a slight modification of the jet distance:
\begin{equation}
d_{ij}^{(flavored)} = d_{ij}^{(standard)} \times
\begin{cases}
\mathcal{S}_{ij} \, , & \text{if both $i$ and $j$ have non-zero flavor of opposite sign,} \\[.2cm] 
1 \, , & \text{otherwise.}
\end{cases}
\label{eq:flavor-anti-kT}
\end{equation}
where
\begin{equation}
 \mathcal{S}_{ij} = 1-\theta\left(1-\kappa_{ij}\right)\cos\left(\frac{\pi}{2}\kappa_{ij}\right)
 \quad \text{with} \quad \kappa_{ij} \equiv \frac{1}{a} \, \frac{k_{T,i}^2+k_{T,j}^2}{2 k_{T,\text{max}}^2}\; .
\label{eq:Sij}
\end{equation}

In our numerical study, we consider the values $a= 0.2, 0.1, 0.05$, which results in the following realizations of the flavored anti-$k_\rT$ algorithm:
 \begin{itemize}
  \item flavored anti-$k_\rT$ algorithm, charge dependent, with $a= 0.2, 0.1, 0.05$ (dubbed a$k_\rT$CD-0.2, a$k_\rT$CD-0.1, and a$k_\rT$CD-0.05, respectively),
  \item flavored anti-$k_\rT$ algorithm, charge agnostic, with $a= 0.1$ (dubbed a$k_\rT$CA-0.1).
 \end{itemize}

Therefore, in total, in this work we consider 7 different flavored jet algorithms.

\subsubsection{Event selection}\label{sec:event-selection}

  For the present work for LHC at $13\TeV$, we take sightly different kinematic cuts than in Ref.~\cite{Czakon:2020coa} where the centre-of-mass energy considered was $7\TeV$.  
  First, the charged lepton ((anti-)muon in our case) has to fulfill the following requirements:
\begin{align}
\label{eq:leptons}
\ptsub{\ell} >  30\GeV, \qquad |\eta_\ell| < 2.5\,.
\end{align}
In addition, at least one c-tagged jet with:
\begin{align}
\label{eq:jet}
 \ptsub{\Pj_c} >  20\GeV, \qquad |\eta_{\Pj_c}| < 2.5 \,,
\end{align}
is required.
At NNLO, in the double real radiation contribution, an event can contain up to three c-jets.

Typically, experimental measurements aim at observing the so-called opposite-sign (OS) contribution which contains a c-jet with electric charge sign opposite to the charge of the charged lepton originating in the $\PW$ decay. This is achieved by removing the same-sign (SS) contribution (which is identified as containing a c-jet and a charged lepton of the same electric charge). The motivation behind the SS/OS denomination is the idea that contributions from $\Pg\to\Pc\bar\Pc$ splittings, which contribute equally to SS and OS but are not directly related to the strange quark content of the proton, are removed. 

The charge of the charm jet is determined by the charge of the lepton resulting from the semileptonic decay of the D meson eventually produced by the fragmenting c-jet. Since our study is performed at the partonic level, the jet charge is determined by the sign of the jet's charm quark and is $+1$ for a $c$ and $-1$ for a $\bar c$.
A comparison to data requires therefore to correct the charm-jet definition with respect to D-meson tagging.
The determination of such corrections is beyond the scope of the present work.
Therefore, we assume that such corrections are provided by the experimental collaborations.

In this work we consider the charge agnostic case with the requirement for at least one c-jet, as well as the following additional event selections
\begin{itemize}
 \item The leading c-jet (based on its transverse momentum) is of OS type, no requirement on c-jet multiplicity,
 \item One and only one c-jet is required, no requirement on c-jet charge,
 \item One and only one c-jet of OS type,
 \item One and only one c-jet of SS type,
 \item OS--SS (``OS {\it minus} SS'') cross section.
\end{itemize}
In all cases, the selection is inclusive in the number of non c-tagged jets.

Finally, for the EW corrections, the radiated photons are recombined with the charged leptons and jets according to the anti-$k_\rT$ algorithm with a radius of $R=0.1$.

\subsection{Tools}

The QCD corrections presented here have  been computed with the help of the {\sc Stripper} program, a \texttt{C++} implementation of the four-dimensional formulation of the sector-improved residue subtraction scheme \cite{Czakon:2010td,Czakon:2011ve,Czakon:2014oma,Czakon:2019tmo}.
With the same code, several V+jets calculations have already been carried out \cite{Czakon:2020coa,Pellen:2021vpi,Pellen:2022fom,Czakon:2022wam,Hartanto:2022qhh,Hartanto:2022ypo}.
The matrix elements have been obtained from the {\sc AvH} library \cite{Bury:2015dla} for tree-level amplitudes and {\sc OpenLoops 2} \cite{Buccioni:2019sur} for the one-loop ones.
On the other hand, the two-loop amplitudes originate from Ref.~\cite{Gehrmann:2011ab} and were numerically evaluated with {\sc Ginac} \cite{Bauer:2000cp,Vollinga:2004sn}.

The NLO EW corrections have been obtained from the private Monte Carlo program {\sc MoCaNLO} in combination with the matrix-element provider {\sc Recola}~\cite{Actis:2012qn,Actis:2016mpe} which has already been used for several V(s)+jets computations \cite{Denner:2019zfp,Biedermann:2016yds,Biedermann:2017bss,Denner:2019tmn,Brauer:2020kfv,Denner:2020zit,Denner:2021hsa,Denner:2022pwc} at NLO EW accuracy.

\section{Theoretical predictions}
\label{sec:th}

In this section, we provide updated predictions for the baseline set-up of Ref.~\cite{Czakon:2020coa} where only events containing at least one c-jet defined with the flavored $k_\rT$ algorithm $k_\rT$CA are kept. Our predictions maintain full CKM dependence through NNLO QCD and utilize the NNPDF3.1 PDF set.

\begin{table}
  \begin{center}
    \begin{tabular}{c|c|c|c}
    Order & $\sigma_{\PW^+\Pj_{\rm c}}$ [$\pb$] & $\sigma_{\PW^-\Pj_{\rm c}}$ [$\pb$] & $R_{\PW^\pm\Pj_{\rm c}} = \sigma_{\PW^+\Pj_{\rm c}}/\sigma_{\PW^-\Pj_{\rm c}}$\\
    \midrule
    LO    & $113.817(2)^{+12.4\%}_{-9.87\%}$      & $119.711(2)^{+12.4\%}_{-9.88\%}$ & $0.95076(2)^{+0.013\%}_{-0.021\%}$ \\
    \midrule
    NLO   & $162.4(1)^{+7.2\%}_{-6.6\%}$        & $168.1(1)^{+6.9\%}_{-6.4\%}$     & $0.9659(9)^{+0.29\%}_{-0.21\%}$ \\
    \midrule
    NNLO  & $168.6(8)^{+0.7\% \; +3.8\% ({\rm PDF})}_{-2.1\% \; -3.8\% ({\rm PDF})}$ & $173.9(1.9)^{+0.6\% \; +3.7\% ({\rm PDF})}_{-1.8\% \; -3.7\% ({\rm PDF})}$ 
    & $0.96(1)^{+0.2\%\; +2.1\% ({\rm PDF})}_{-0.3\%\; -2.1\% ({\rm PDF})}$ \\
    \end{tabular}
  \end{center}
  \caption{\label{tab:XsectionQCD}
    Fiducial cross sections for $\Pp\Pp \to \PW^+\Pj_{\rm c}$, $\Pp\Pp \to \PW^-\Pj_{\rm c}$, and their ratios at the LHC at $\sqrt{s}=13\TeV$ at LO, NLO, and NNLO \QCD.
    The digit in parenthesis indicates the Monte Carlo statistical error while the sub- and super-script in per cent indicate the scale variation.
    In addition, the PDF variation is provided for the NNLO QCD predictions.
    The full CKM matrix and the NNLO NNPDF3.1 set with $\alphas=0.118$ are used for all predictions. The c-jets are defined with the $k_\rT$CA algorithm with the at least one c-jet requirement.}
\end{table}

In Table~\ref{tab:XsectionQCD}, the fiducial cross section is given at LO, NLO, and NNLO QCD accuracy.
The QCD corrections show good perturbative convergence.
In particular, NNLO QCD corrections are significantly smaller than the NLO ones.
Note that, at variance with Ref.~\cite{Czakon:2020coa} and following the PDF4LHC recommendations~\cite{Butterworth:2015oua,Ball:2022hsh}, NNLO PDF sets are used for all predictions at all orders.
While the centre-of-mass energy is different from the one used in Ref.~\cite{Czakon:2020coa}, the smaller corrections can principally be explained with the different choice of PDF at LO and NLO accuracy.
We also note that the NNLO $K$-factor in the ratio $R_{\PW^\pm\Pj_{\rm c}}$ is essentially 1.
It implies that this ratio constitutes a particularly reliable observable as it is rather stable under perturbative corrections.
As already pointed out in Ref.~\cite{Czakon:2020coa}, the PDF uncertainty reaches almost $4\%$ and is larger than the scale uncertainty at NNLO QCD which varies between $0.6\%$ and $2.1\%$. The inclusion of NNLO QCD corrections therefore allows for a clean future determination of the strange-quark content of the proton from this observable. 

\begin{table}
  \begin{center}
    \begin{tabular}{c|c|c|c}
    Order & $\sigma_{\PW^+\Pj_{\rm c}}$ [$\pb$] & $\sigma_{\PW^-\Pj_{\rm c}}$ [$\pb$] & $R_{\PW^\pm\Pj_{\rm c}} = \sigma_{\PW^+\Pj_{\rm c}}/\sigma_{\PW^-\Pj_{\rm c}}$\\
    \midrule
    NLO EW & $117.399(2)$      & $111.627(2)$ & $0.95084(2)$ \\
    \midrule
    \midrule
    $\delta_{\rm NLO\; EW} [\%]$   & $-1.93$        & $-1.92$     & $-0.01$ \\
    \end{tabular}
  \end{center}
  \caption{\label{tab:XsectionEW}
    Fiducial cross sections and relative NLO EW corrections at order $\mathcal{O} \left(\alphas \alpha^3 \right)$ for $\Pp\Pp \to \PW^+\Pj_{\rm c}$, $\Pp\Pp \to \PW^-\Pj_{\rm c}$, and their ratios at the LHC at $\sqrt{s}=13\TeV$.
    No QCD corrections are included in these predictions.
    The digit in parenthesis indicates the Monte Carlo statistical error.
    The full CKM matrix and the NNLO NNPDF3.1 set with $\alphas=0.118$ are used.}
\end{table}

In addition, in Table~\ref{tab:XsectionEW}, the NLO EW corrections are provided for both signatures.
It is interesting to notice that both processes receive almost the same corrections, resulting therefore in an almost zero correction at the level of the $R_{\PW^\pm\Pj_{\rm c}}$ ratio.
This is not a surprise given that at the LHC, EW corrections are largely driven by Sudakov logarithms~\cite{Denner:2019vbn}.
The latter depend on the quantum numbers of the external states and the typical scale of the process~\cite{Denner:2000jv}.
Given that for both signatures the quantum numbers are identical and the typical scales are very close, the corrections are almost exactly the same.
This implies that the corrections essentially do not contribute at the level of the ratio (tenth of a per mille), therefore reinforcing the statement made above that this ratio is very stable under perturbative corrections in the Standard Model.

\begin{figure}
        \begin{subfigure}{0.49\textwidth}
                 \includegraphics[width=\textwidth,page=1]{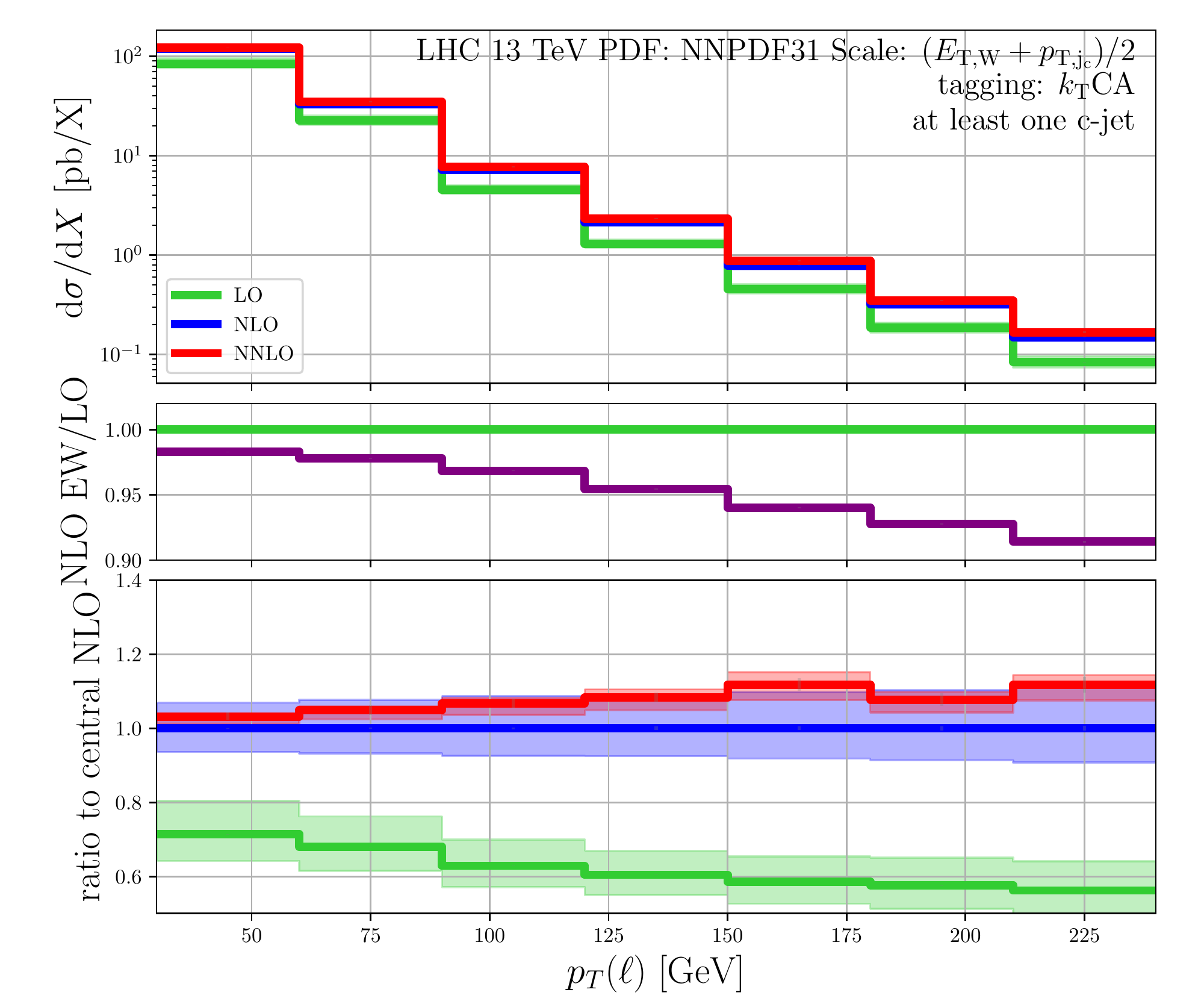}
        \end{subfigure}
        \hfill
        \begin{subfigure}{0.49\textwidth}
                 \includegraphics[width=\textwidth,page=3]{plots/best_predictions/pQCDEW_ktv2_incl_NNPDF31_muWJ}
        \end{subfigure}

        \caption{\label{fig:bestpred}%
                Differential distributions in the transverse momentum (left) and the absolute rapidity (right) of the charged lepton for the process $\Pp\Pp \to \PW^+\Pj_{\rm c}$ at $\sqrt{s}=13\TeV$.
                The upper panel shows the LO, NLO, and NNLO QCD absolute predictions without EW corrections.
                The middle panel represents the NLO EW corrections normalised to the LO predictions.
                The lower panel displays the LO and NNLO QCD predictions and data relative to the NLO QCD prediction.
                }
\end{figure}

In Fig.~\ref{fig:bestpred}, differential distributions in the transverse momentum and the absolute rapidity of the charged lepton are shown for the plus signature.
We refrain from showing results for the minus signature as they are qualitatively very similar.
As observed at the level of the cross section and in Ref.~\cite{Czakon:2020coa}, the QCD corrections are characterised by large NLO $K$-factors and moderate NNLO ones.
As usual, these corrections are accompanied by a significant reduction of the theoretical uncertainty estimated via scale variation.
This observation holds for both observables as well as for the transverse momentum and rapidity of the charm jet (not shown).

On the other hand, the EW corrections have a rather different behaviour for the two observables.
For the transverse momentum, the corrections become negative and large when going towards higher energy.
At low transverse momentum, the corrections are at the level of few per cent, as for the fiducial cross section, while they reach almost $10\%$ above $200\GeV$.
This behaviour is typical for EW corrections which are driven by Sudakov logarithms in the high-energy limit.
The situation is rather different for the rapidity of the charged lepton.
In this case, there is no enhancement due to Sudakov logarithms at high energy.
The corrections are thus flat and the offset is inherited from the corrections to the fiducial cross section.

It is worth pointing out that the rapidity distribution of the charged lepton also receives moderate QCD corrections with moderate shape distortion.
It means that the rapidity distribution of the charged lepton is largely insensitive to higher-order corrections of both QCD and EW type, making it therefore a perfect observable to be measured experimentally.

\section{Detailed analysis}\label{sec:details}

\subsection{Off-diagonal CKM elements}

In Ref.~\cite{Czakon:2020coa}, our best predictions at NNLO QCD accuracy included the effects of non-zero $V_{\Pc\Pd}$ element at LO only.
In that previous work, we anticipated the effect to be ``probably within few per cent'' with respect to a full computation with off-diagonal CKM elements.
Table~\ref{tab:CKM} confirms this expectation.
It provides NNLO QCD predictions with full off-diagonal CKM dependence (dubbed \emph{full CKM}), with only $V_{\Pc\Pd}\neq0$ at LO (dubbed \emph{$V^{\rm LO}_{\Pc\Pd}\neq0$}), and with no off-diagonal dependence (dubbed \emph{no CKM}).
For the plus signature and for the minus signature the differences amount to about $2.5\%$ and $3\%$, respectively.
On the other hand, for the two signatures, not considering any off-diagonal CKM elements amounts to an effect of roughly $7\%$ and $10\%$, respectively.

\begin{table}
  \begin{center}
    \begin{tabular}{c|c|c|c}
     $\sigma_{\rm NNLO}$ [$\pb$] & full CKM  & $V^{\rm LO}_{\Pc\Pd}\neq0$ & no CKM \\
    \midrule
    + & $168.6(8)^{+0.7\% \; +3.8\% ({\rm PDF})}_{-2.1\% \; -3.8\% ({\rm PDF})}$ & $164.4(8)^{+1.0\% \; +3.9\% ({\rm PDF})}_{-2.4\% \; -3.9\% ({\rm PDF})}$ & $156.7(8)^{+0.7\% \; +4.2\% ({\rm PDF})}_{-2.1\% \; -4.2\% ({\rm PDF})}$ \\
    \midrule
    - & $173.9(1.9)^{+0.6\% \; +3.7\% ({\rm PDF})}_{-1.8\% \; -3.7\% ({\rm PDF})}$  & $168.5(1.9)^{+1.0\% \; +3.8\% ({\rm PDF})}_{-2.2\% \; -3.8\% ({\rm PDF})}$ & $156.7(1.9)^{+0.5\% \; +4.2\% ({\rm PDF})}_{-1.6\% \; -4.2\% ({\rm PDF})}$ \\
    \end{tabular}
  \end{center}
  \caption{\label{tab:CKM}
    Fiducial cross sections with full off-diagonal CKM dependence (\emph{full CKM}), with only $V_{\Pc\Pd}\neq0$ at LO (\emph{$V^{\rm LO}_{\Pc\Pd}\neq0$}), and with no off-diagonal dependence (\emph{no CKM}).
    All predictions are at NNLO QCD accuracy and are shown for both the $\Pp\Pp \to \PW^+\Pj_{\rm c}$ and $\Pp\Pp \to \PW^-\Pj_{\rm c}$ process at $\sqrt{s}=13\TeV$.
    The digit in parenthesis indicates the Monte Carlo statistical error while the sub- and super-script in per cent indicate the scale variation.
    In addition, the PDF variation is also provided for the NNLO NNPDF3.1 set with $\alphas=0.118$.}
\end{table}

\begin{figure}
        \begin{subfigure}{0.49\textwidth}
                 \includegraphics[width=\textwidth,page=1]{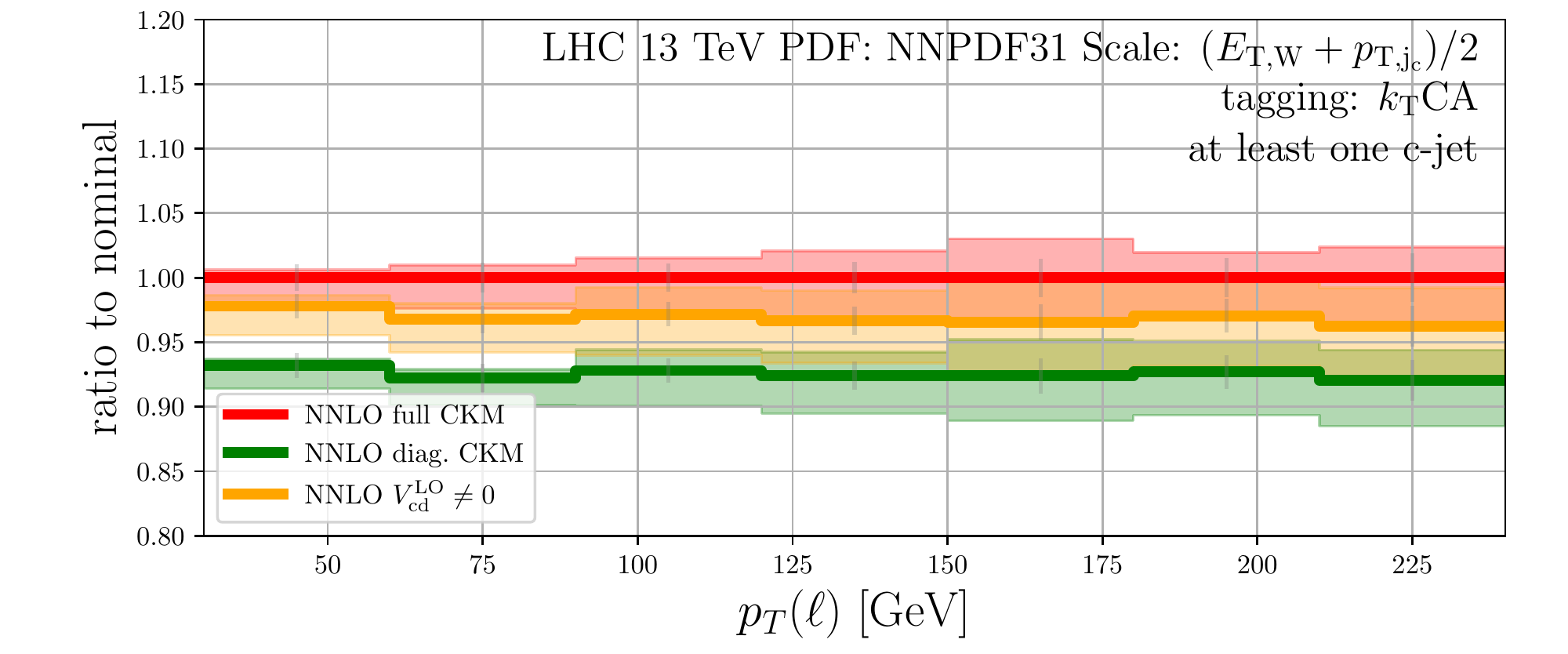}
        \end{subfigure}
        \hfill
        \begin{subfigure}{0.49\textwidth}
                 \includegraphics[width=\textwidth,page=2]{plots/best_predictions/CKM_NNLO_incl_NNPDF31_muWJ}
        \end{subfigure}

        \caption{\label{fig:CKM}%
                Ratios of differential distributions in the transverse momentum of the charged lepton (left) and the charm jet (right) for the process $\Pp\Pp \to \PW^+\Pj_{\rm c}$ at $\sqrt{s}=13\TeV$.
                It shows the NNLO QCD predictions including only diagonal CKM elements (green) and the $V_{\Pc\Pd}$ element included at LO only (orange),  normalised to the prediction with all off-diagonal CKM elements included (red).
                }
\end{figure}

This observation can also be made for the differential distributions in Fig.~\ref{fig:CKM} which show the \emph{$V^{\rm LO}_{\Pc\Pd}\neq0$} and \emph{no CKM} predictions normalised by the \emph{full CKM} ones for the transverse momentum of the charged lepton and the charm jet.
The ratio plots do not show significant shape distortion and the $K$-factor is largely inherited from the differences observed at the level of the fiducial cross section.
We note that while the predictions have significant statistical uncertainties with respect to the ratio, one can still make a reliable statements about the differences of the various predictions as they are statistically fully correlated, i.e. they are based on the same sample of phase space points.
The same holds true for the rest of the article when ratio plots are displayed.

\subsection{PDF dependence and scale setting}

In this section, the dependence of the predictions on the PDF set and scale choice is discussed.
In particular, the PDF-set choice is of crucial importance due to the sensitiviy to the strange-quark content of the proton.

\begin{figure}
        \begin{subfigure}{0.49\textwidth}
                 \includegraphics[width=\textwidth,page=1]{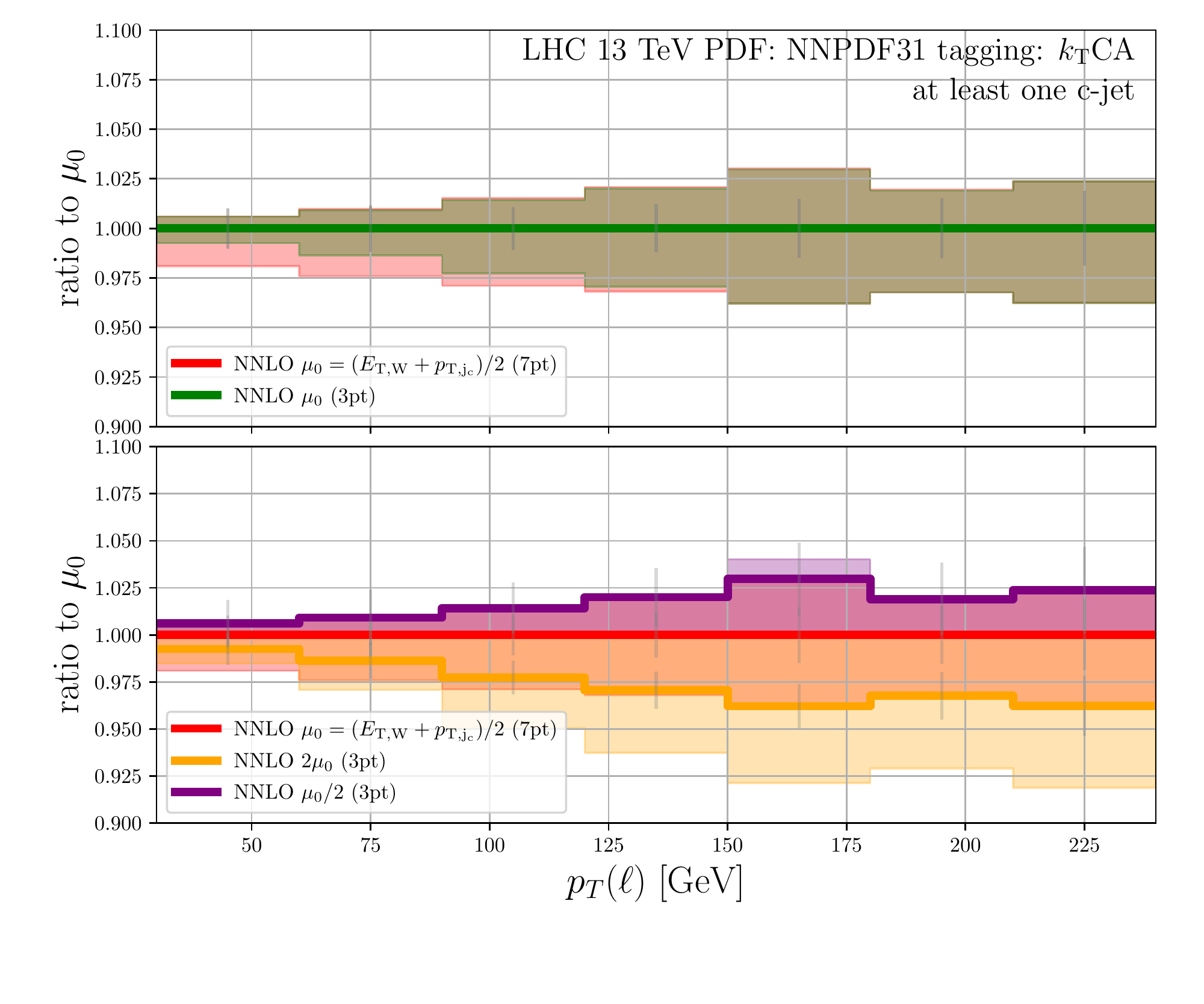}
        \end{subfigure}
        \hfill
        \begin{subfigure}{0.49\textwidth}
                 \includegraphics[width=\textwidth,page=1]{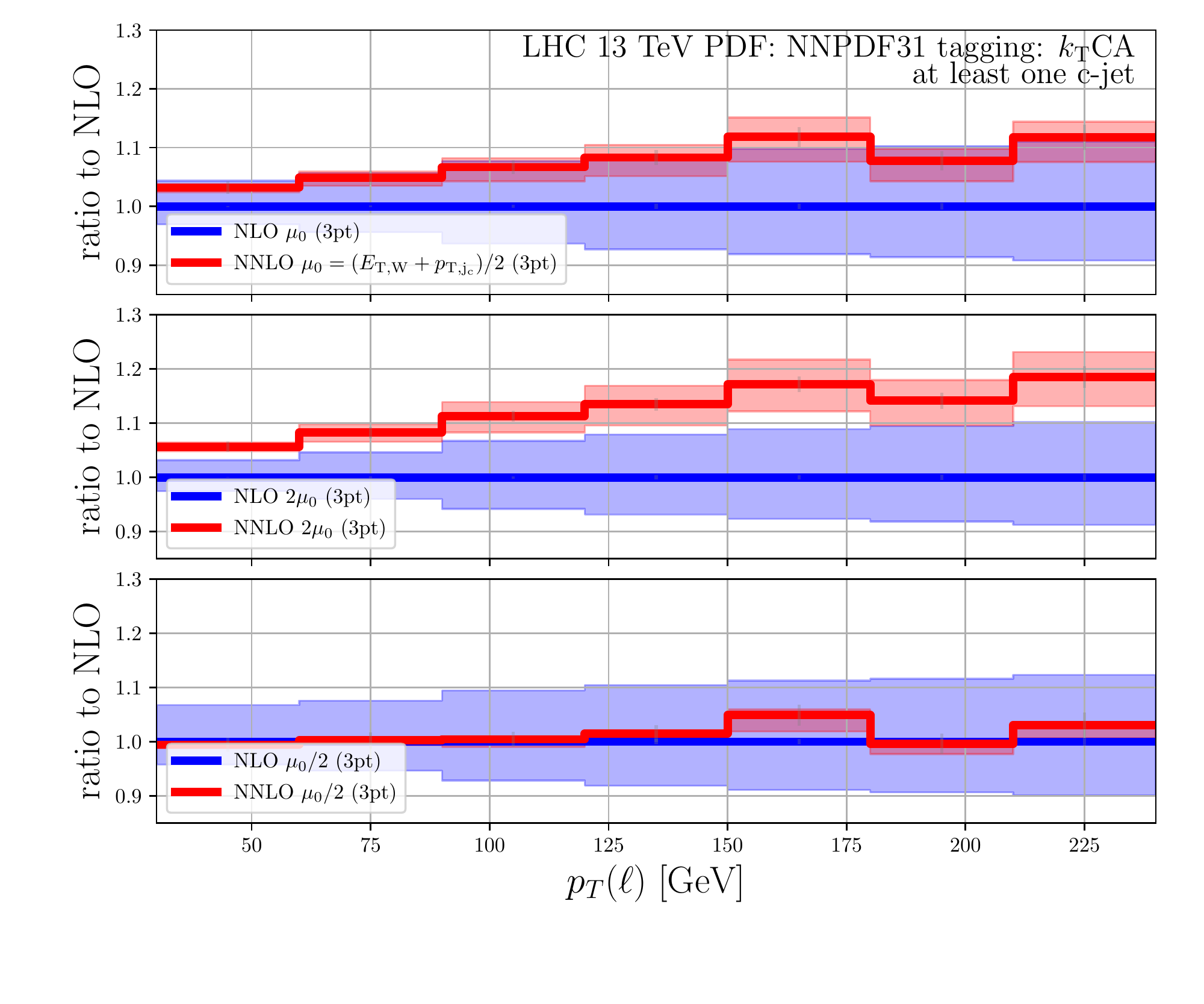}
        \end{subfigure}

        \caption{\label{fig:scale}%
                Ratios of differential distributions in the transverse momentum of the charged lepton for the process $\Pp\Pp \to \PW^+\Pj_{\rm c}$ at $\sqrt{s}=13\TeV$.
                Left: A comparison of 7-points and 3-points scale variation for the nominal scale (top) and a comparison of different central values with 3-points scale variation (bottom). Right: A comparison of NNLO $K$-factors for different central scales with 3-points scale variation.
                }
\end{figure}

First, in Fig.~\ref{fig:scale} the transverse momentum of the charged lepton is shown for different central scales and different prescriptions for the scale variation.
On the left-hand side, the 7-points and 3-points scale-variation prescriptions are compared (upper plot).
For the transverse momentum, they agree rather well at high-transverse momentum while at low transverse momentum, the 7-points variation is larger.
Nonetheless, the differences do not exceed $2\%$.
In the lower part, three different central scales are compared:
the nominal one $\mu_0$ from Eq.~\eqref{eq:scale} as well as half and twice this scale.
At low transverse momentum, the three choices agree within $3\%$ while at $250\GeV$, the spread reaches more than $5\%$.
On the right-hand side of Fig.~\ref{fig:scale}, NNLO $K$-factors are shown for the three different scales.
For this observable, the smallest corrections are obtained for $\mu_0/2$.
On the other hand, for the transverse momentum of the hardest c-jet (not shown), the smallest corrections are obtained for $\mu_0$.
In general, at the level of the fiducial cross section, the smallest cross section is obtained with $2\mu_0$.
Note that for the plots with the 3 different scales, the 3-points scale variation prescription is used.

\begin{figure}
        \begin{subfigure}{0.49\textwidth}
                 \includegraphics[width=\textwidth,page=1]{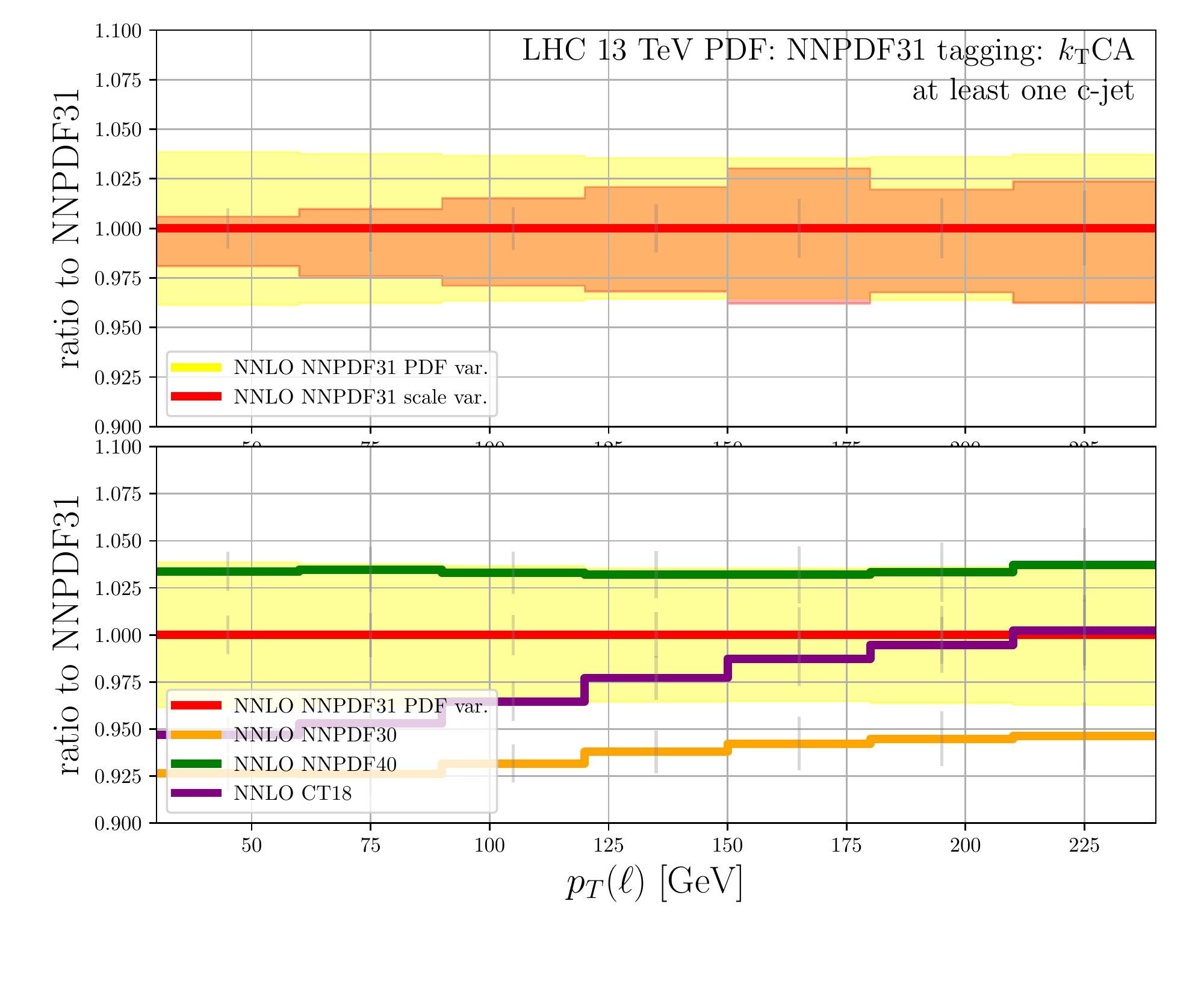}
        \end{subfigure}
        \hfill
        \begin{subfigure}{0.49\textwidth}
                 \includegraphics[width=\textwidth,page=3]{plots/variations/variations_pdf_NNLO_incl_muWJ_ktv2.pdf}
        \end{subfigure}

        \caption{\label{fig:PDF}%
                Ratio of differential distributions in the transverse momentum of the charged lepton (left) and the rapidity of the charged lepton (right) for the process $\Pp\Pp \to \PW^+\Pj_{\rm c}$ at $\sqrt{s}=13\TeV$.
                The top panels show a comparison of 7-points scale variation against PDF error. The lower panels show the PDF error for the NNPDF3.1 set compared to the central predictions of the NNPDF3.0, NNPDF4.0, and CT18 sets.
                }
\end{figure}

In Fig.~\ref{fig:PDF}, the transverse momentum and the rapidity distributions of the charged lepton are shown.
In the upper plots, the 7-points scale variation is compared to the PDF uncertainty.
As at the level of the cross section (see Table~\ref{tab:XsectionQCD}) and at $7\TeV$~\cite{Czakon:2020coa}, NNLO scale variation is smaller than the PDF uncertainty.
This implies that NNLO predictions are crucial for the precise determination of strange and anti-strange PDFs.
In particular, NLO QCD predictions are insufficient for constraining PDFs given that the NLO scale uncertainty is about twice the PDF uncertainty (see Table~\ref{tab:XsectionQCD}).
In the lower plots, the predictions are provided at NNLO QCD for different sets.
In addition to the nominal one (NNPDF3.1), we also show predictions for the NNPDF3.0, NNPDF4.0, and CT18 sets.
It is worth emphasising that there are large variations between the different sets.
While the predictions with NNPDF4.0 are within the PDF uncertainty of NNPDF3.1, this is not the case for CT18 across the whole phase space.
Interestingly, the predictions obtained with the NNPDF3.0 set are always outside of the PDF uncertainty band of the NNPDF3.1 set.

\begin{table}
  \begin{center}
    \begin{tabular}{c|c|c|c|c}
    $\sigma_{\rm NNLO}$ [$\pb$] & NNPDF3.1 & NNPDF4.0 & NNPDF3.0 & CT18 \\
    \midrule
    + & $168.6(8)^{+0.7\% \; +3.8\% ({\rm PDF})}_{-2.1\% \; -3.8\% ({\rm PDF})}$ & $174.3(8)$ & $156.3(8)$ & $160.1(8)$ \\
    \midrule
    - & $173.9(1.9)^{+0.6\% \; +3.7\% ({\rm PDF})}_{-1.8\% \; -3.7\% ({\rm PDF})}$ & $183.7(1.9)$ & $160.3(1.7)$ & $163.0(1.7)$ \\
    \end{tabular}
  \end{center}
  \caption{\label{tab:pdf}
    Fiducial cross sections with full off-diagonal CKM dependence at NNLO QCD accuracy for both the $\Pp\Pp \to \PW^+\Pj_{\rm c}$ and $\Pp\Pp \to \PW^-\Pj_{\rm c}$ process at $\sqrt{s}=13\TeV$.
    The predictions are provided for four different PDF sets:
    NNPDF3.1 (default), NNPDF4.0, NNPDF3.0, and CT18.
    The digit in parenthesis indicates the Monte Carlo statistical error.
    The scale variation and the PDF variation is provided for the NNPDF3.1 set only.}
\end{table}

For completeness, we also provide in Table~\ref{tab:pdf} the central values for the four different PDF sets.
As at the differential level, we can observe that the spread is of the order of $10\%$.
In particular, among all the theoretical effects that we study in details in the present work, the PDF is the largest source of uncertainty.
This therefore strongly motivates the effort for improving the determination of strange PDFs using state-of-the-art theory predictions presented in this work.

\subsection{Event selection and same-sign contribution}
\label{sec:selection}

In this section, we scrutinise various event selections related to the definition of the charm jet.
Experimental measurements usually provide the OS--SS cross section.
As explained in sec.~\ref{sec:event-selection}, the motivation behind this fact is to get rid of contributions of the type $qq'\to\PW+(\Pg\to\Pc\bar\Pc)$ which do not carry a dependence on the strange-quark PDF. Since such channels contribute equally to the OS and SS cross sections they are excluded from the OS--SS cross section.

\begin{table}
  \begin{center}
    \begin{tabular}{c|c|c|c|c|c}
    $\sigma$ [$\pb$] & incl. & leading c-jet OS & exactly one c-jet & exactly one OS c-jet & OS--SS\\
    \midrule
    $\sigma^+_{\rm NLO}$ & $162.4(1)$ & $156.9(1)$ & $161.0(1)$ & $156.1(1)$ & $151.1(1)$ \\
    \midrule
    $\sigma^-_{\rm NLO}$ & $168.1(1)$ & $164.0(1)$ & $166.9(1)$ & $163.3(1)$ & $159.7(1)$ \\
    \midrule
    \midrule
    $\sigma^+_{\rm NNLO}$ & $168.6(8)$ & $159.0(8)$ & $165.8(8)$ & $157.3(8)$ & $148.9(8)$ \\
    \midrule
    $\sigma^-_{\rm NNLO}$ & $173.9(1.9)$ & $166.8(1.8)$ & $171.5(1.9)$ & $165.2(1.8)$ & $159.0(1.7)$ \\
    \end{tabular}
  \end{center}
  \caption{\label{tab:select}
    Fiducial cross sections at NLO QCD and NNLO QCD accuracy for different charm jet selections: at least one c-jet (\emph{incl.}), leading c-jet is OS, exactly one c-jet, exactly one OS c-jet, and the OS--SS selection.
    The digit in parenthesis indicates the Monte Carlo statistical error.}
\end{table}

In the following we show predictions for the selections specified in sec.~\ref{sec:event-selection}.
The various cross sections at NLO and NNLO QCD are tabulated in Table~\ref{tab:select}.
Note that at LO, there is no dependence on the c-jet selection given that there is only one parton in the final state.

As expected, the highest cross section corresponds to the selection with at least one c-jet as it is inclusive in the charm jets.
The second highest cross section corresponds to exactly one c-jet.
The third in size is obtained by enforcing the leading c-jet to be of OS type which is very close to the requirement for only one c-jet of OS type.
The lowest cross sections, the OS--SS one, differs from the exactly one OS c-jet selection by the size of the SS contribution.
It is interesting to notice that all these definitions differ by at most $5\%$ at NLO QCD accuracy while they can differ by almost $10\%$ at NNLO QCD.
This is due to the fact that NNLO corrections contain double-real effects with $\Pc\Pc\bar\Pc$ or $\Pc\bar\Pc\bar\Pc$ final states.
Note that the selection choice in Ref.~\cite{Czakon:2020coa} was to retain events with one and only one c-jet.

It is interesting to note that in Ref.~\cite{Bevilacqua:2021ovq}, where NLO QCD predictions with parton shower (PS) corrections were computed, 
the size of the SS contribution at $7\TeV$ has been found to be between $5\%$ and $10\%$ for the $\PW+{\rm D}$-meson and $\PW+{\rm D^*}$-meson signatures while it is slightly below $3\%$ for the $\PW+\Pj_{\rm c}$ final state.
For the charm-jet final state, we find a similar order of magnitude at NLO QCD accuracy.
On the other hand, the SS contributions can grow to about $5\%$ due to the double-real contributions at NNLO.

 \begin{figure}
        \begin{subfigure}{0.49\textwidth}
                 \includegraphics[width=\textwidth,page=3]{plots/selection/CjetVeto_ktv2_NLO_NNPDF31_muWJ}
        \end{subfigure}
        \hfill
        \begin{subfigure}{0.49\textwidth}
                 \includegraphics[width=\textwidth,page=1]{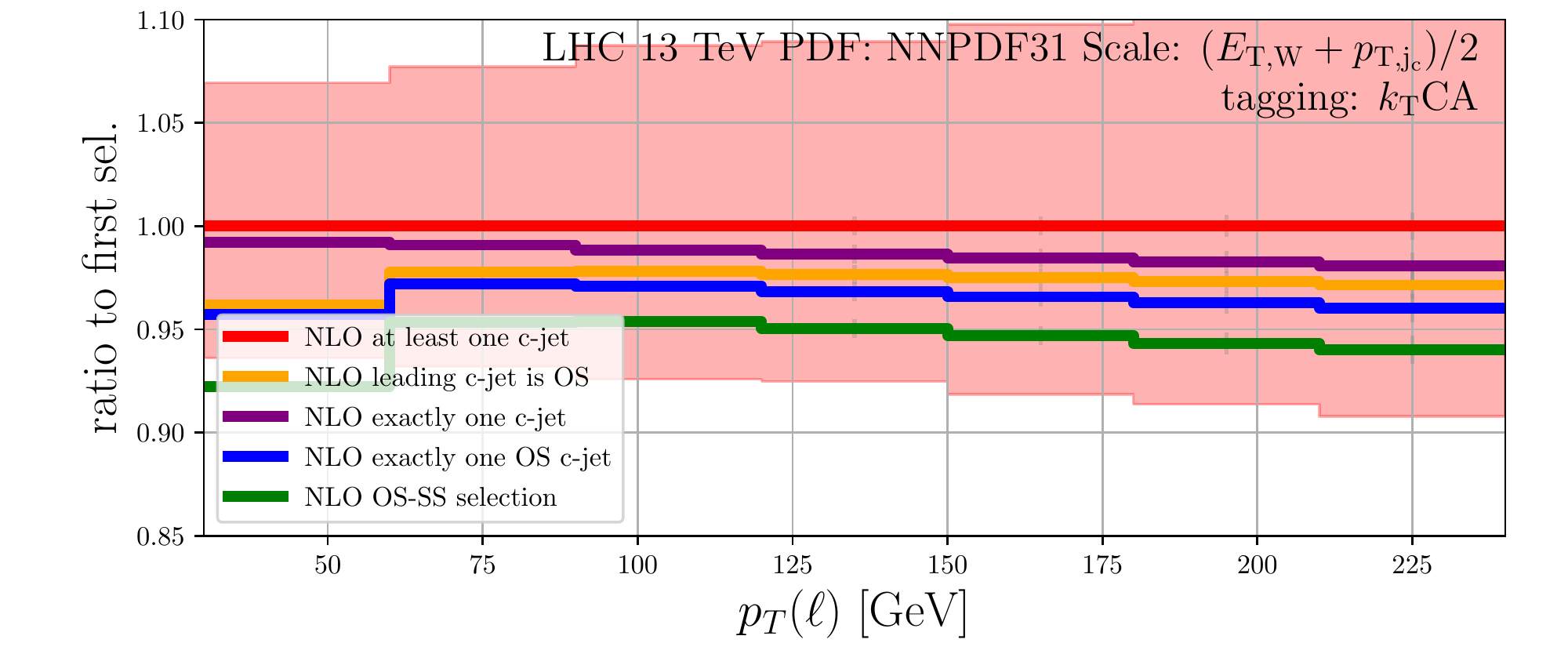}
        \end{subfigure}

        \begin{subfigure}{0.49\textwidth}
                 \includegraphics[width=\textwidth,page=3]{plots/selection/CjetVeto_ktv2_NNLO_NNPDF31_muWJ}
        \end{subfigure}
        \hfill
        \begin{subfigure}{0.49\textwidth}
                 \includegraphics[width=\textwidth,page=1]{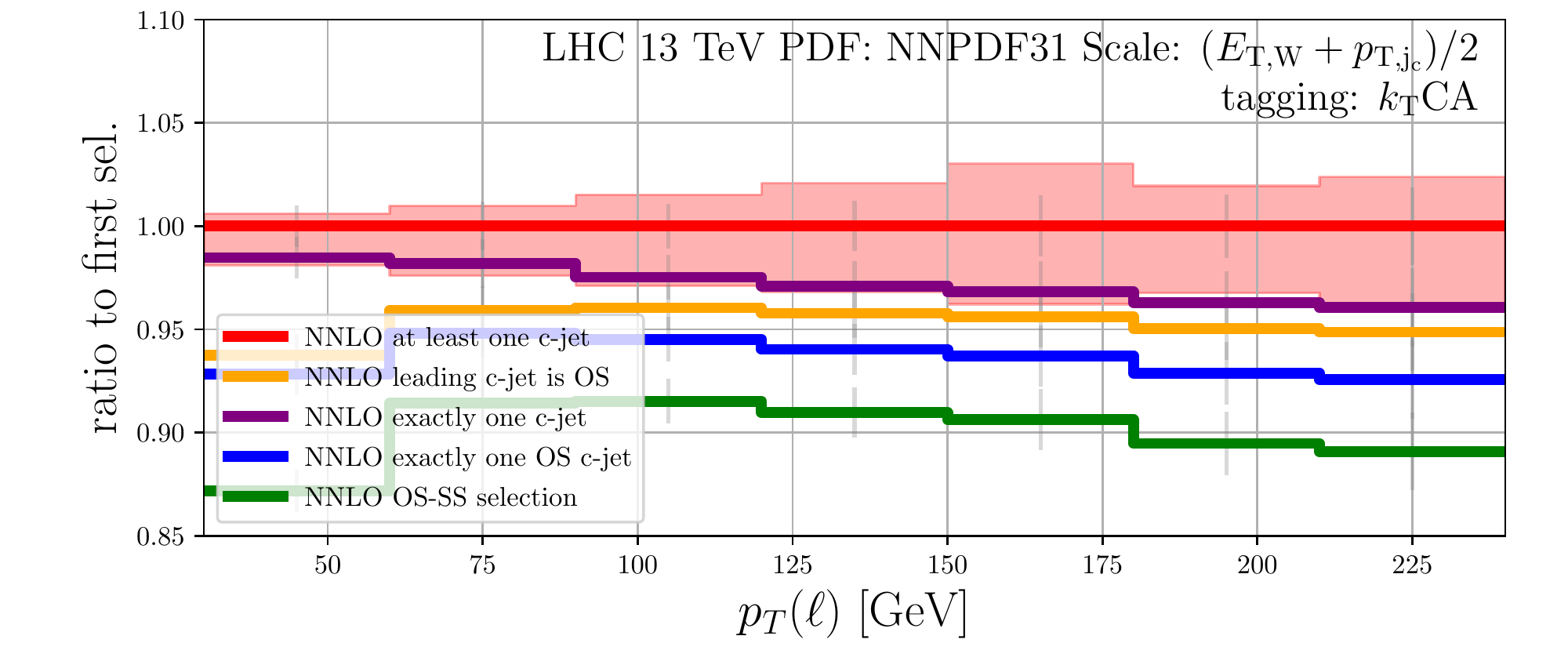}
        \end{subfigure}
        
        \caption{\label{fig:select}%
                Ratio of differential distributions in the rapidity (left) and the transverse momentum (right) of the charged lepton for the process $\Pp\Pp \to \PW^+\Pj_{\rm c}$ at $\sqrt{s}=13\TeV$.
                The upper plots show the results at NLO QCD while the lower ones are for NNLO QCD.
                Various event selection for the charm jet are compared: at least one c-jet, the leading c-jet being of OS type, exactly one c-jet, exactly one c-jet of OS type, and the OS--SS selection.}
\end{figure}

Fig.~\ref{fig:select} provides the same information as Table~\ref{tab:select} but differentially, in terms of the rapidity and transverse momentum of the charged lepton.
For the rapidity of the charged lepton, the differences between the various selections are most significant in the high-rapidity region.
This is explained by the fact that the SS contribution is larger in this region.
The size of the SS contribution can be inferred by comparing the selections with exactly one c-jet and with exactly one OS c-jet.
At NLO, the difference is about $2.5\%$ at zero rapidity and it reaches $5\%$ at the highest rapidity $y(\ell)=2.5$.
This contribution is enhanced at NNLO QCD with difference of about $5\%$ and $
7\%$ at low and large rapidity, respectively.
Regarding the transverse momentum distribution, one observes shape differences between the various selections.
The differences are maximal at low transverse momentum (below $50\GeV$), about $7\%$ at NLO and $13\%$ at NNLO QCD.
The differences are smallest around $75\GeV$ and start to increase again towards large transverse momenta at both NLO and NNLO.

Finally, recall that the motivation for the OS--SS cross section is its direct link with the strange-quark PDF. This relation is based on a LO argument which is modified once off-diagonal CKM elements and higher-order corrections are included. Such effects dilute the sensitivity of this selection to the strange quark content of the proton and must be carefully accounted for in any precision extraction of the strange quark PDF.

\subsection{Flavor jet algorithms}

This section is devoted to the comparison of various jet algorithms that are used for defining the process under study.
We first focus on the differences between various $k_\rT$ algorithms, after which we consider a newly-introduced flavored anti-$k_\rT$ algorithm.
Finally, flavored algorithms are compared against the standard anti-$k_\rT$ algorithm for NLO QCD+PS predictions.

\subsubsection{Flavor $k_\rT$ algorithms}\label{sec:flavor-kT-algos}

\begin{figure}
        \begin{subfigure}{0.49\textwidth}
                 \includegraphics[width=\textwidth,page=3]{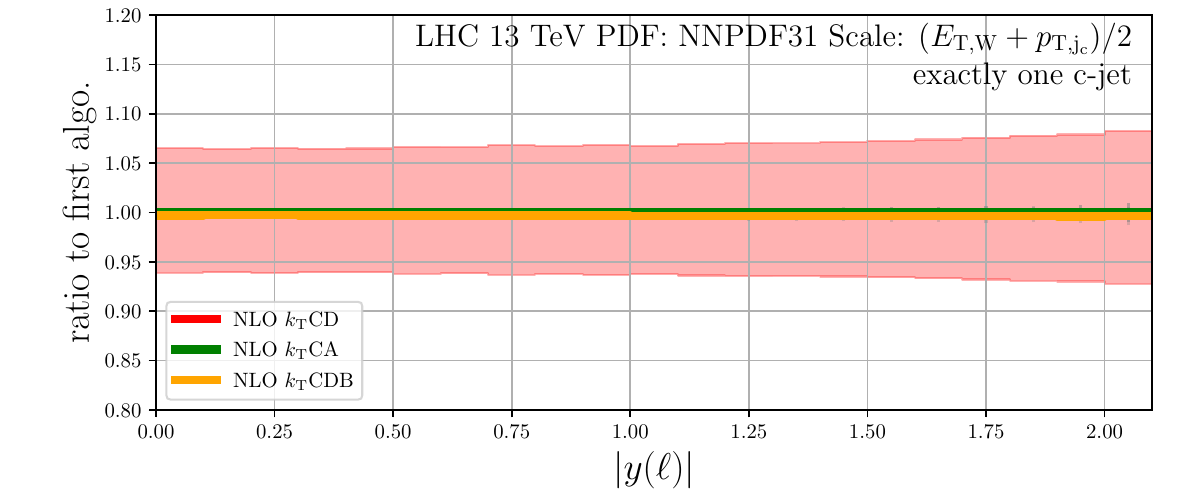}
        \end{subfigure}
        \hfill
        \begin{subfigure}{0.49\textwidth}
                 \includegraphics[width=\textwidth,page=3]{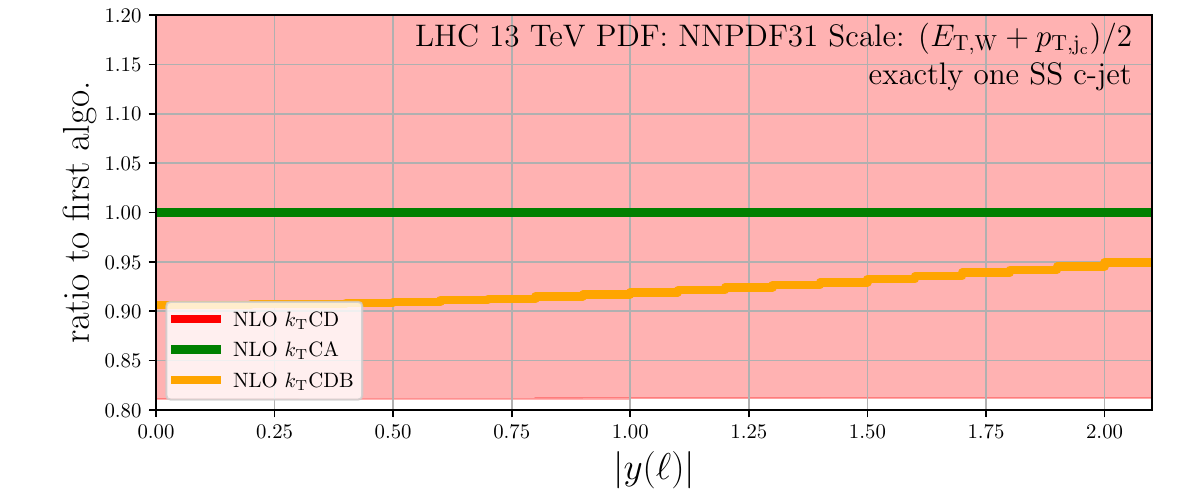}
        \end{subfigure}

        \begin{subfigure}{0.49\textwidth}
                 \includegraphics[width=\textwidth,page=3]{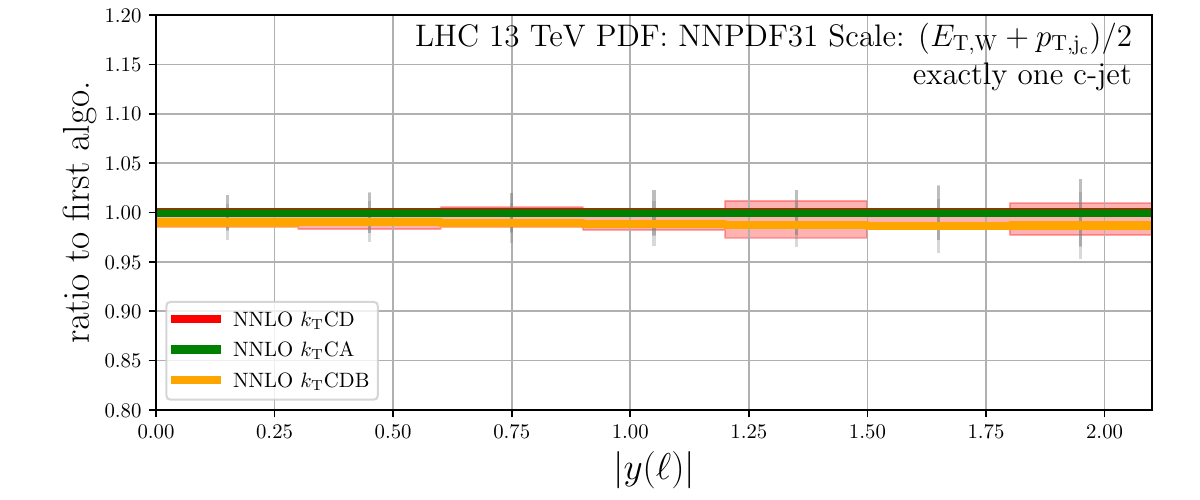}
        \end{subfigure}
        \hfill
        \begin{subfigure}{0.49\textwidth}
                 \includegraphics[width=\textwidth,page=3]{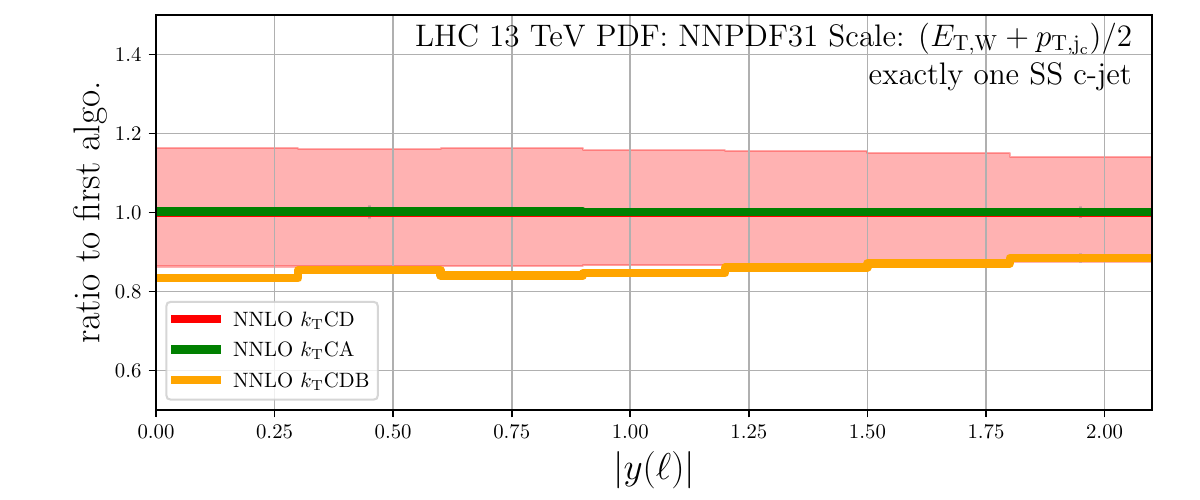}
        \end{subfigure}
        
        \caption{\label{fig:flkt}%
                Ratio of differential distributions in the rapidity of the charged lepton for the process $\Pp\Pp \to \PW^+\Pj_{\rm c}$ at $\sqrt{s}=13\TeV$.
                The upper plots show the results at NLO QCD while the lower ones are for NNLO QCD.
                The left-hand side plots are for one and only c-jet while the right-hand side ones are for one and only one c-jet of SS type.
                Various definition of the flavored $k_\rT$ algorithm are compared (see text).
                }
\end{figure}

In this subsection, we compare the three implementations of the flavored $k_\rT$ algorithm \cite{Banfi:2006hf} listed in sec.~\ref{sec:jet-algos}. Their comparison for the absolute rapidity of the charged lepton, at NLO and NNLO QCD, is shown in Fig.~\ref{fig:flkt}. Results are presented for the two event selections given in sec.~\ref{sec:event-selection}, namely, one and only one c-jet and one and only one c-jet of SS type.
For the charge dependent c-jet selection, it is interesting to observe that the differences in the implementation of the flavored $k_\rT$ algorithms have little effect on the differential results.
In particular, the differences are well within the scale uncertainty band.
This conclusion holds at both NLO and NNLO QCD accuracy as well as for other observables like the transverse momentum of the charged lepton, the transverse momentum of the charm jet or the charm-jet rapidity (not shown).

The situation is rather different when selecting only one c-jet of SS type.
While choosing the jet algorithm to be either charge agnostic or charge dependent has no effect, including the W momenta in the beam definition of the algorithm has a large effect.
At NLO QCD, the effects are about $10\%$ at zero rapidity and $5\%$ at maximal rapidity.
At NNLO QCD, the effects are even more significant, reaching more than $15\%$ for central rapidities and more than $10\%$ in the peripheral region.
The same level of differences can be observed in other differential distributions.

It is particularly interesting to notice that while the SS contribution shows a large dependence on the algorithm definition, this dependence is essentially absent when not specifying the charge of the charm jet.
This is simply due to the fact that the SS contribution is rather small with respect to the OS one at $13\TeV$ (see Section~\ref{sec:selection}).
Therefore, the large differences disappear when adding SS and OS cross sections in a charge-agnostic selection.

\subsubsection{Flavor anti-$k_\rT$ algorithms}\label{sec:flavor-anti-kT-algos}

In this subsection, we consider the implementations of the flavored anti-$k_\rT$ algorithm \cite{Czakon:2022wam} specified in sec.~\ref{sec:jet-algos}. All the variants are compared against the flavored $k_\rT$ algorithm $k_\rT$CD which is charge dependent.

 \begin{figure}
        \begin{subfigure}{0.49\textwidth}
                 \includegraphics[width=\textwidth,page=1]{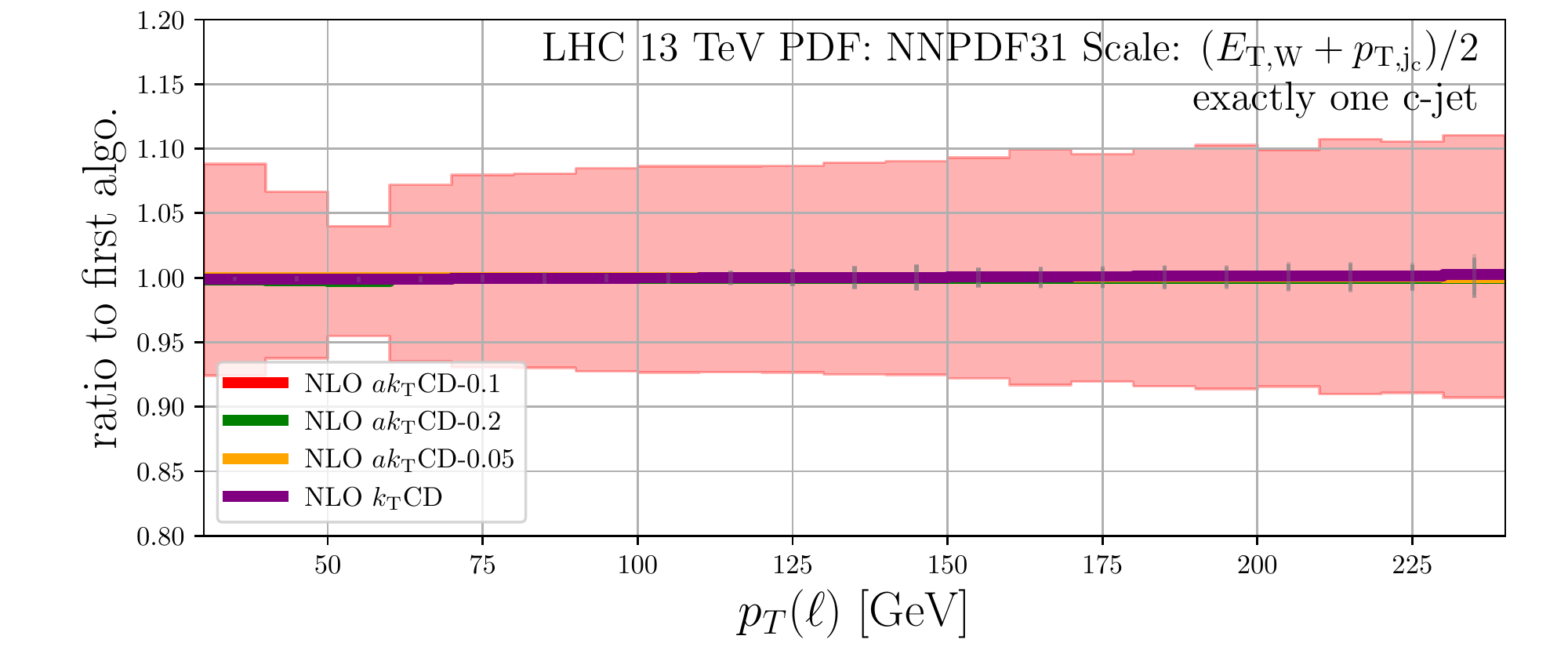}
        \end{subfigure}
        \hfill
        \begin{subfigure}{0.49\textwidth}
                 \includegraphics[width=\textwidth,page=1]{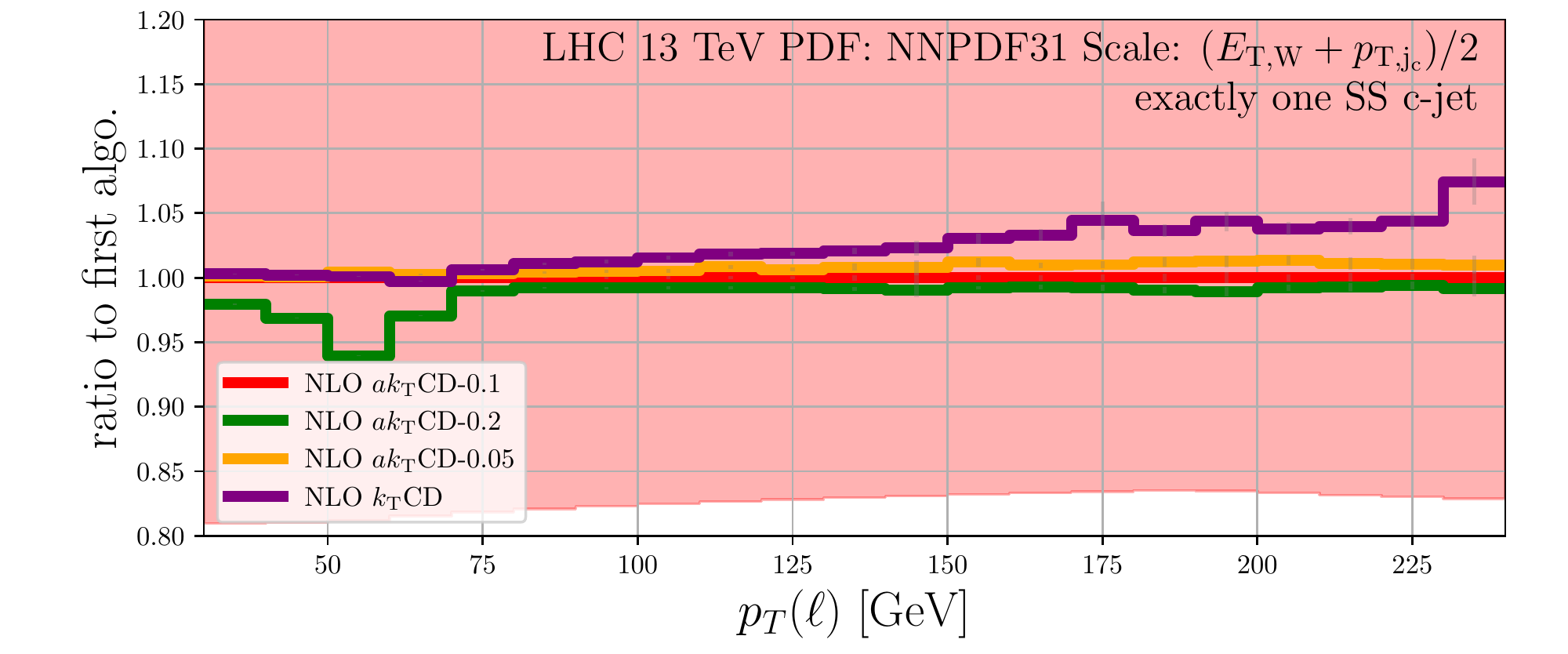}
        \end{subfigure}

        \begin{subfigure}{0.49\textwidth}
                 \includegraphics[width=\textwidth,page=1]{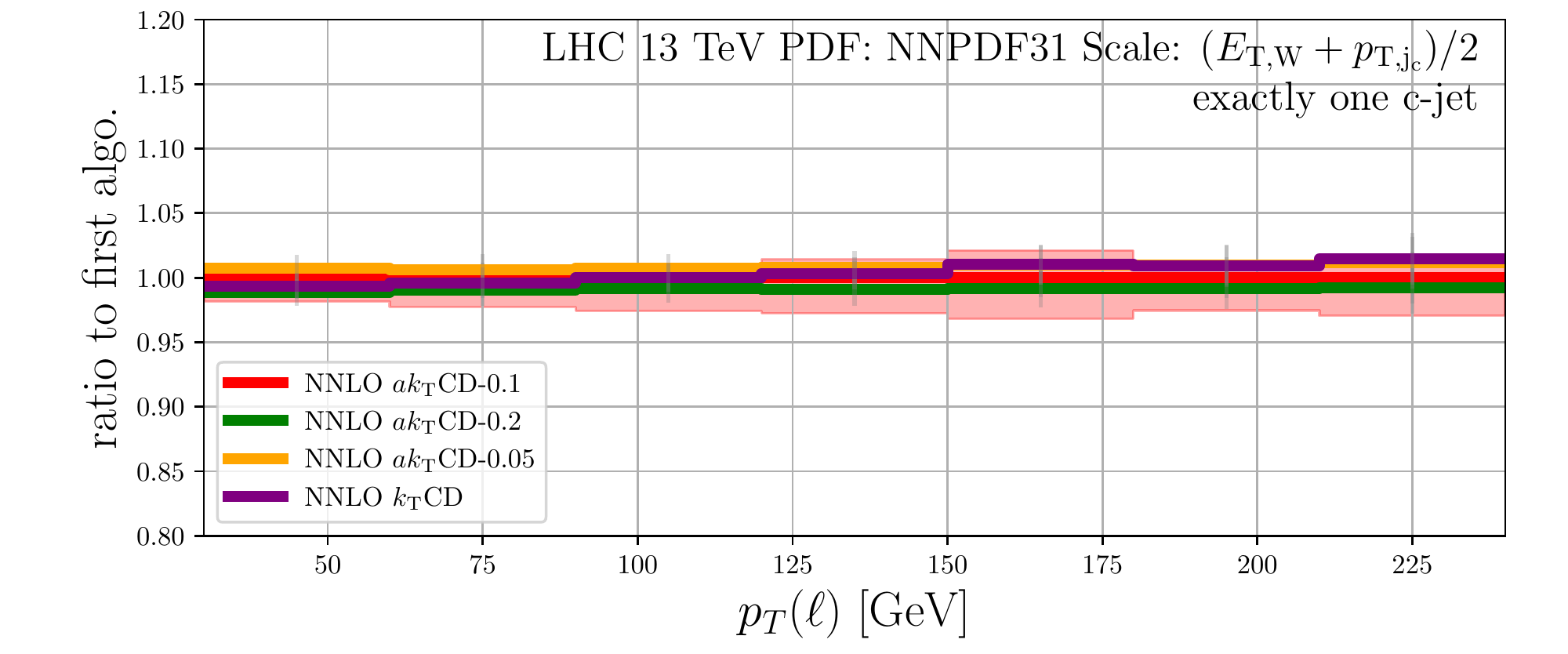}
        \end{subfigure}
        \hfill
        \begin{subfigure}{0.49\textwidth}
                 \includegraphics[width=\textwidth,page=1]{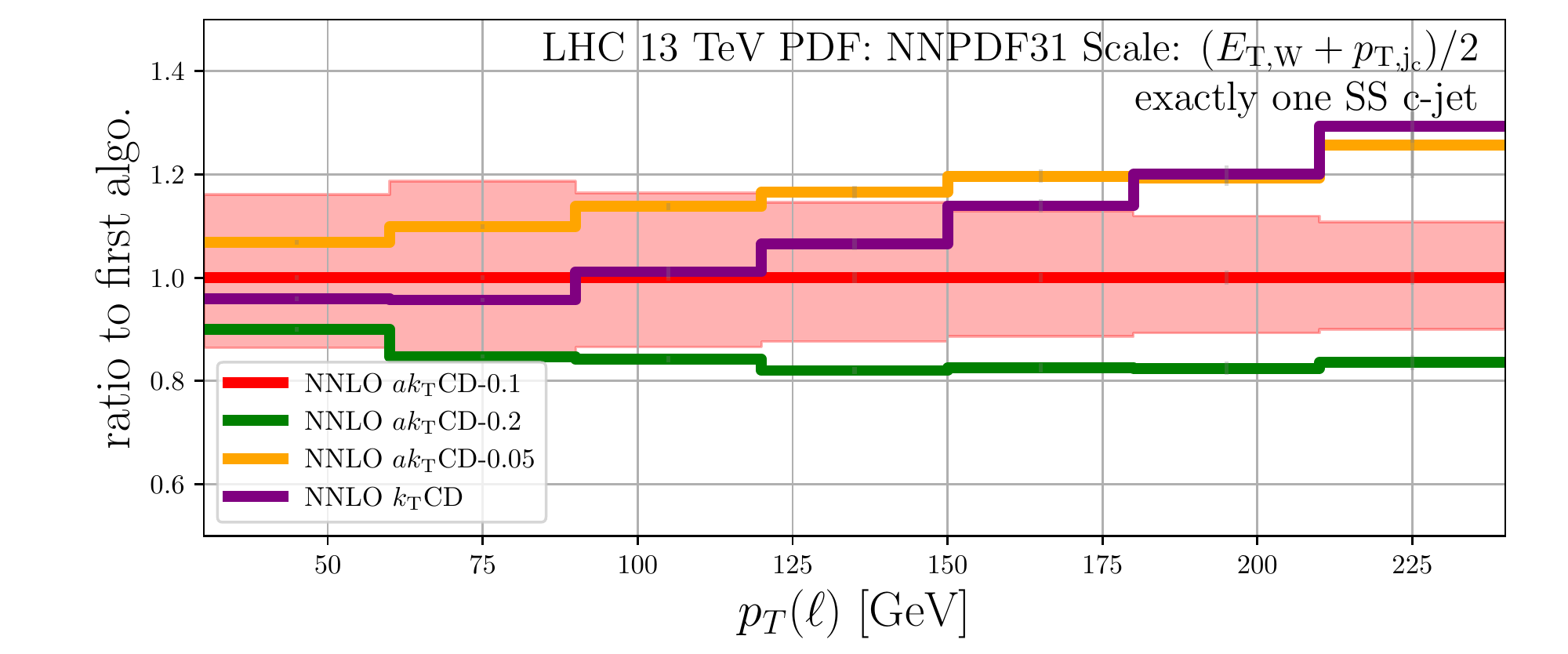}
        \end{subfigure}
        
        \caption{\label{fig:flakt}%
                Ratio of differential distributions in the rapidity of the charged lepton for the process $\Pp\Pp \to \PW^+\Pj_{\rm c}$ at $\sqrt{s}=13\TeV$.
                The upper plots show the results at NLO QCD while the lower ones are for NNLO QCD.
                The left-hand side plots are for one and only c-jet while the right-hand side ones are for one and only one c-jet of same-sign type.
                Various definition of the flavored anti-$k_\rT$ algorithm are compared.
                In addition, these predictions are compared to the nominal flavored $k_\rT$ algorithm (see text).
                }
\end{figure}

In Fig.~\ref{fig:flakt} we consider the transverse momentum of the charged lepton computed with the flavored anti-$k_\rT$ algorithm for different values of the $a$ parameter. Since the value of this parameter is not set from first principle, we vary it in the range between $0.2$ and $0.05$. We observe that, essentially, this variation has no impact on the predictions for the exactly-one-c--jet selection at both NLO and NNLO QCD. Furthermore, all these predictions are within the scale variation band and are also in perfect agreement with the nominal $k_\rT$ algorithm.
This holds true not only for the transverse momentum of the charged lepton but also for other standard observables like the $p_T$ and rapidity of $\Pj_{\rm c}$ and the charge muon's rapidity.

This situation is in stark contrast with the selection containing only one SS c-jet. In this case, at NLO in QCD, the flavored anti-$k_\rT$ algorithm with $a=0.2$ differs from the ones with $a=0.1, 0.05$ by about $5\%$ around $60\GeV$. At high transverse momentum the nominal flavored $k_\rT$ algorithm differs from the flavored anti-$k_\rT$ algorithm with $a=0.1$ by about $5\%$. These differences become more pronounced at NNLO in QCD where the three anti-$k_\rT$ algorithms almost never agree within scale uncertainty, their differences ranging between $5\%$ and $20\%$. Similarly, the transverse momentum of the charged lepton shows a completely different behaviour between the nominal flavored $k_\rT$ algorithm and the flavored anti-$k_\rT$ algorithm with $a=0.1$. As can be seen in Fig.~\ref{fig:flakt}, the difference between the two algorithms becomes larger than $20\%$ at high transverse momentum (above $200\GeV$). 

Unlike the case of NLO QCD, at NNLO QCD the scale uncertainty band does not cover these differences. This behaviour is analogous to the one already discussed in sec.~\ref{sec:flavor-kT-algos} for the flavored $k_\rT$ algorithms.

\subsubsection{NLO QCD with parton shower corrections}

A suitable value for the parameter $a$ entering the flavored anti-$k_\rT$ algorithm was determined in ref.~\cite{Czakon:2022wam} based on the idea that predictions from the standard anti-$k_\rT$ and the flavored anti-$k_\rT$ algorithms are close. Due to the lack of flavored IR-safety for the standard anti-$k_\rT$ algorithm, such a comparison can only be done at NLO with the help of a parton shower. In this section we extend the study of ref.~\cite{Czakon:2022wam} to the present context of W+c production. Such a study is also interesting given the large sensitivity of SS events to the value of the anti-$k_\rT$ algorithm's $a$-parameter, see the discussion in sec.~\ref{sec:flavor-anti-kT-algos}.

In this section we consider the transverse momentum and rapidity of the charged lepton. The results are obtained with the help of \madgraph~\cite{Alwall:2014hca} at NLO QCD matched to the {\sc Pythia} parton shower~\cite{Bierlich:2022pfr} with default parameters. Note that while all input parameters and event selections are identical to the ones used for the fixed-order results, the renormalization and factorization scales are chosen to be $H_{\rm T}$, the default scale choice in \madgraph\footnote{For this reason, we refrain from comparing the NLO QCD+PS predictions against the fixed-order ones.}.
The events have been written into \textsc{Hepmc} files \cite{Dobbs:2001ck} and are analysed using the \textsc{Rivet} analysis framework \cite{Bierlich:2019rhm}.

\begin{figure}
        \begin{subfigure}{0.49\textwidth}
                 \includegraphics[width=\textwidth,page=1]{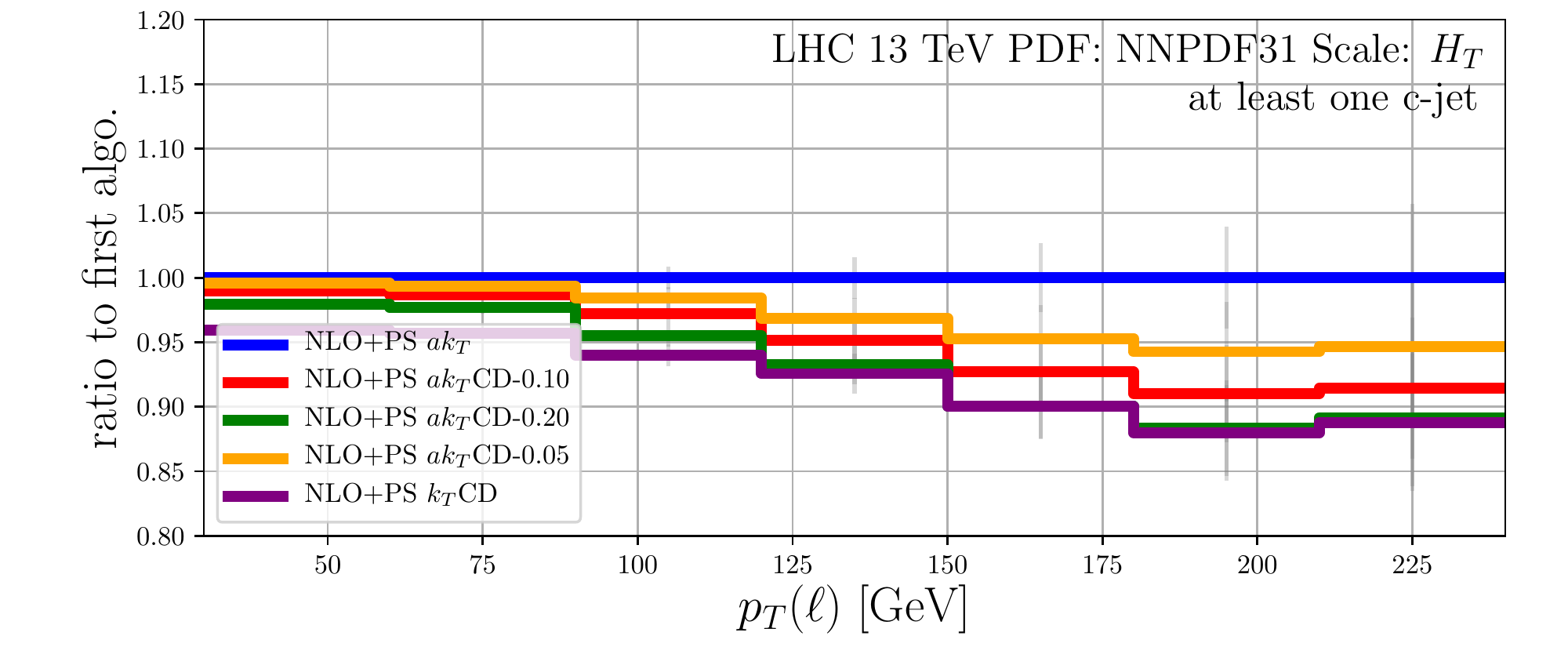}
        \end{subfigure}
        \hfill
        \begin{subfigure}{0.49\textwidth}
                 \includegraphics[width=\textwidth,page=2]{plots/nlops/ppwc_algorithms_NLO+PS}
        \end{subfigure}

        \begin{subfigure}{0.49\textwidth}
                 \includegraphics[width=\textwidth,page=11]{plots/nlops/ppwc_algorithms_NLO+PS}
        \end{subfigure}
        \hfill
        \begin{subfigure}{0.49\textwidth}
                 \includegraphics[width=\textwidth,page=12]{plots/nlops/ppwc_algorithms_NLO+PS}
        \end{subfigure}
        
        \caption{\label{fig:algoNLOPS}%
                Differential distributions in the transverse momentum (top) and the rapidity of the charged lepton (bottom) for the process $\Pp\Pp \to \PW^+\Pj_{\rm c}$ at $\sqrt{s}=13\TeV$.
                All results are at NLO QCD+PS accuracy.
                The left-hand side plots are for charge agnostic c-jet while the right-hand side ones are for the leading jet being of OS type.
                Various definitions of the flavored anti-$k_\rT$ algorithm are compared.
                In addition, these predictions are compared to the nominal flavored $k_\rT$ algorithm as well as the standard anti-$k_\rT$ algorithm. Vertical bars show statistical uncertainty.
                }
\end{figure}
\begin{figure}
        \begin{subfigure}{0.49\textwidth}
                 \includegraphics[width=\textwidth,page=4]{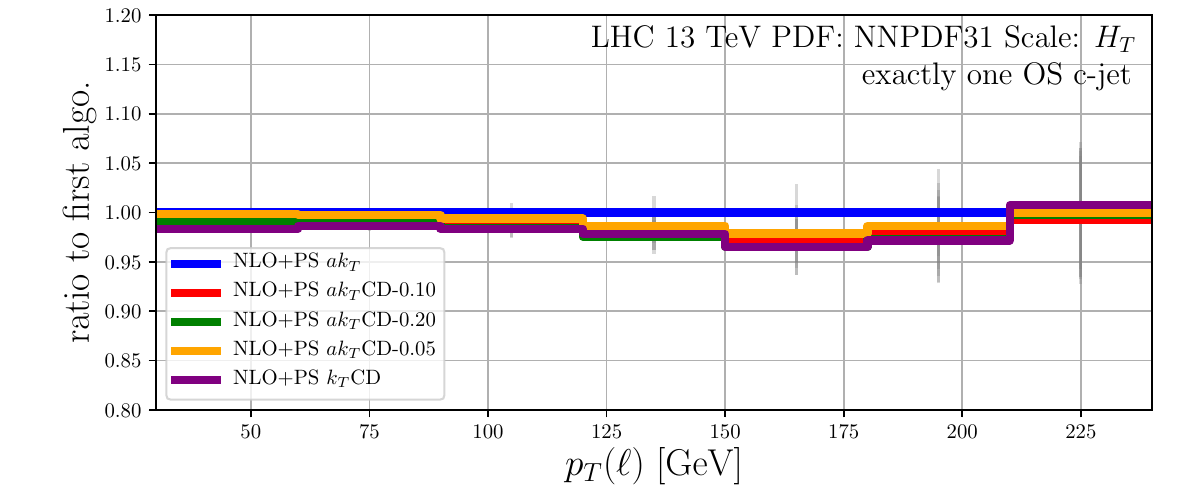}
        \end{subfigure}
        \hfill
        \begin{subfigure}{0.49\textwidth}
                 \includegraphics[width=\textwidth,page=5]{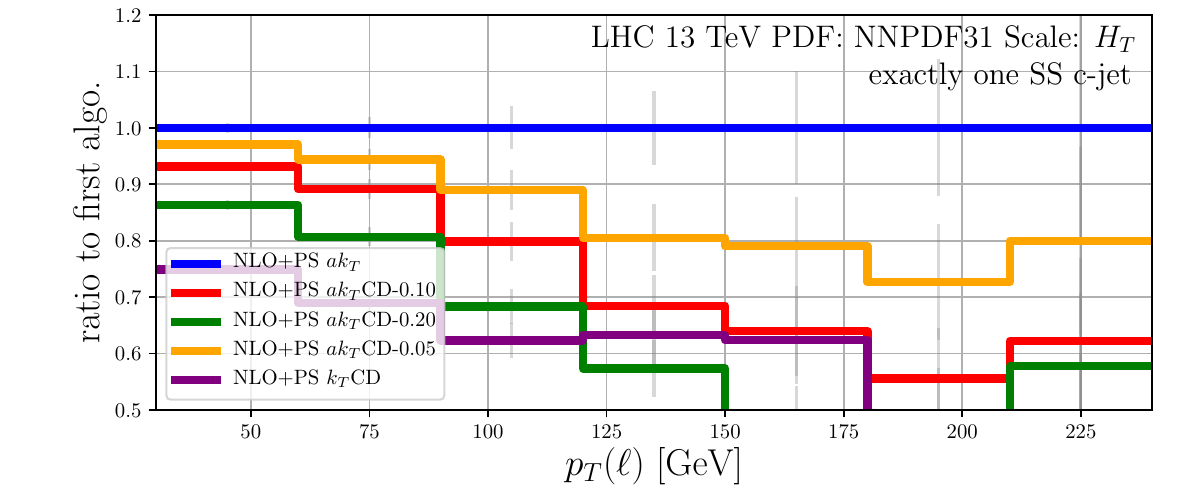}
        \end{subfigure}

        \begin{subfigure}{0.49\textwidth}
                 \includegraphics[width=\textwidth,page=14]{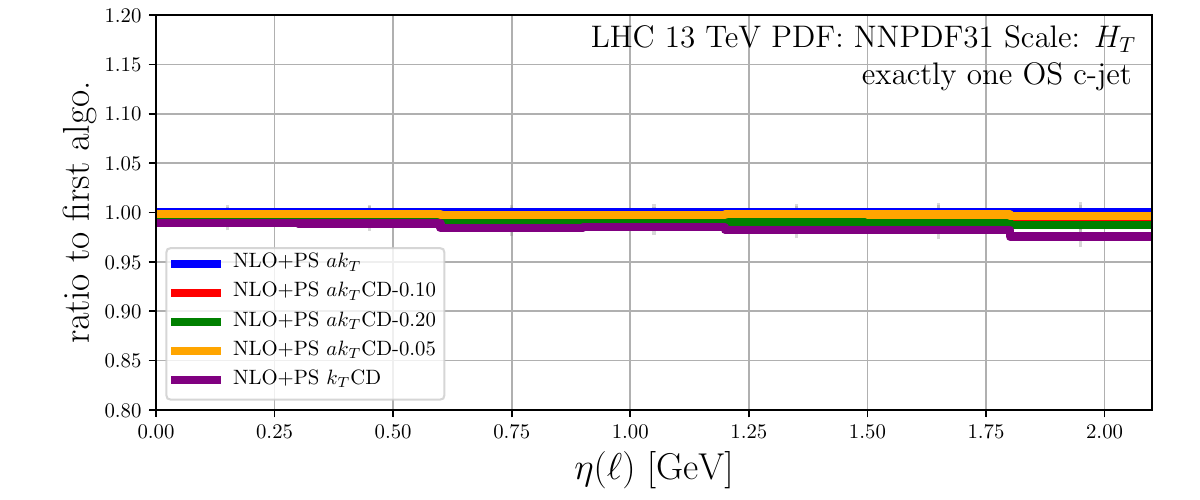}
        \end{subfigure}
        \hfill
        \begin{subfigure}{0.49\textwidth}
                 \includegraphics[width=\textwidth,page=15]{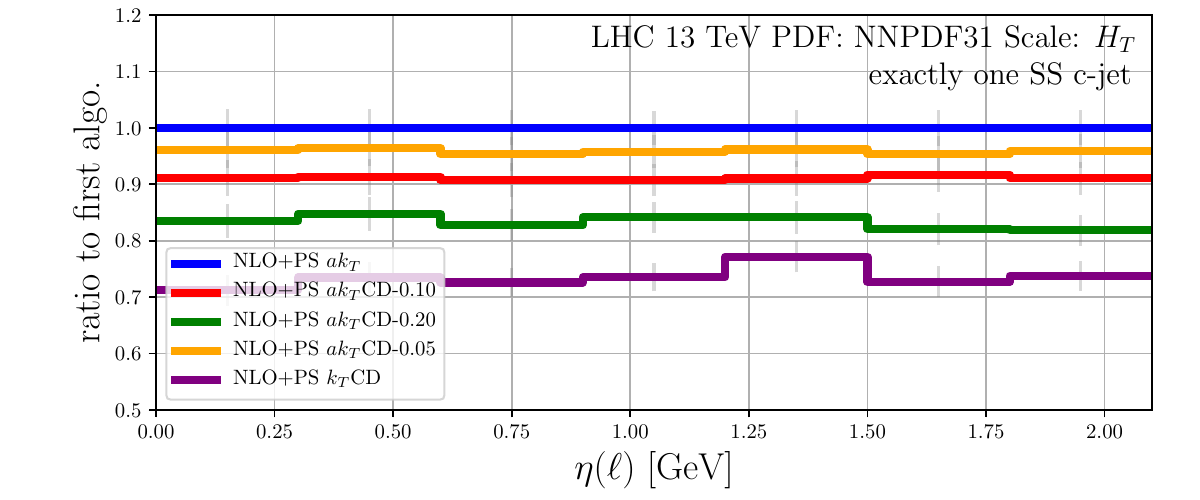}
        \end{subfigure}
        
        \caption{\label{fig:algoNLOPS2}%
                As in fig.~\ref{fig:algoNLOPS} but showing exactly one OS c-jet (left) and exactly one SS c-jet (right).}
\end{figure}

In Fig.~\ref{fig:algoNLOPS}, results for the transverse momentum (top) and the rapidity (bottom) of the charged lepton for different algorithms are shown for a charge agnostic selection of the charm jets (left) as well as requiring that the leading c-jet is of OS type (right).
Again, to be concise, the present results are for the plus signature.

For the transverse-momentum distribution where there are no requirement on the sign of the charm jet (top-left), we observe a spread of about $10\%$ between the different jet algorithms at about $200\GeV$. Nonetheless, for the bulk of the cross section \emph{i.e.}\ below $100\GeV$, the differences do not exceed $5\%$.
In particular, the difference between the anti-$k_\rT$ algorithm and the flavored anti-$k_\rT$ variant with $a=0.1$ (our nominal choice) is below $2\%$.
For the rapidity distribution, we do not observe noticeable shape differences over the full range.
The difference between the anti-$k_\rT$ algorithm and the flavored anti-$k_\rT$ variant with $a=0.1$ is also around $1-2\%$.

On the right-hand side of Fig.~\ref{fig:algoNLOPS}, a different selection is used, namely that the leading charm jet is of OS type.
One observes qualitative similarities with the results for the charge agnostic c-jet selection: in the low transverse-momentum region, and over the whole rapidity range, the differences between the anti-$k_\rT$ algorithm and the nominal flavored anti-$k_\rT$ variant with $a=0.1$ are around $1\%$. The reason for the small difference between these two selections is, as explained previously, in the smallness of the SS contribution.

In Fig.~\ref{fig:algoNLOPS2} (left) we show the results for a selection where exactly one OS c-jet is present. The predictions are similar to the one for the selection where the leading c-jet is OS however the dependence on the jet definition gets significantly reduced. The main difference between these two selections is that the former one is less likely to contain $\Pc\bar\Pc$ pairs in the final state. Clearly, this comparison independently confirms the observation that the increase in jet definition sensitivity is related to the presence of $\Pc\bar\Pc$ pairs. Indeed, the same observation can readily be made for the SS c-jet selection shown in Fig.~\ref{fig:algoNLOPS2} (right). This selection is dominated by $\Pc\bar\Pc$ pairs and just as observed in the NNLO case in sec.~\ref{sec:flavor-anti-kT-algos}, shows very strong sensitivity to the jet algorithm also at NLO+PS. 

Overall, with the nominal choice of flavored anti-$k_\rT$ ($a=0.1$), the difference for charge-agnostic or OS selections with what is, to a good approximation, done in experimental analyses is small. As a reference, this difference is comparable to the size of the missing higher-order corrections of QCD type at NNLO QCD and is significantly smaller than the PDF uncertainty.

\section{Conclusions}
\label{sec:concl}

In this article, we perform a detailed theoretical investigation of W+c-jet production at the LHC. Extending our previous work~\cite{Czakon:2020coa}, we address several open questions, for example the size of off-diagonal CKM contributions beyond NLO QCD, the size of NLO EW corrections, PDF uncertainties, the effect of charm-jet selections, and finally, the effect of flavored jet algorithms. We also provide state-of-the art predictions for present and future W+c-jet measurements at the LHC.

Electroweak corrections at NLO are found to be at the level of $-1.9\%$ for the fiducial cross section and do not contribute in the ratio of the two signatures. Still, they can approach $-10\%$ in differential distributions in the high-energy limit (for example above $200\GeV$ for the transverse momentum of the charged lepton).

The size of the non-diagonal CKM contributions at NNLO QCD is of the order of $10\%$. We have checked that the simple-to-implement approximation where $V_{\Pc\Pd}\neq0$ contributions are included at LO only (as was done in Ref.~\cite{Czakon:2020coa}), already agrees with the full result within $3\%$. Still a 3\% effect is comparable to, and often larger than, the size of the scale uncertainty at NNLO and is, therefore, consequential in any precision study of W+charm.

At $13\TeV$, just like for LHC at $7\TeV$ \cite{Czakon:2020coa}, the scale uncertainty of NNLO QCD is significantly smaller than the PDF uncertainty. By comparing fiducial cross section predictions based on different PDF sets, we observe a spread between the different predictions that can be as large as $10\%$. This strong PDF dependence can be viewed as an opportunity for the precision extraction of the strange quark PDF from LHC data. 

The differences between the charge agnostic, OS and OS--SS charm-jet selections at the differential distribution level can be up to $15\%$ for the kinematics ranges considered in this work. We find that these selections exhibit little-to-mild sensitivity (of up to few per cent) to the parametrization of the flavored jet algorithm. On the other hand, the so-called SS selection exhibits strong sensitivity to the details of the jet algorithm and is numerically much smaller - at the level of 5\% - than the other charm jet selections. This behaviour of the SS cross section is to be expected since it is the one predominantly containing $\Pc\bar\Pc$ pairs in the final state. While our findings are specific to the process we study (W+c) some lessons might translate to other processes, like Z+c. In particular, in Z+c the partonic channels with gluon splittings to $\Pc\bar\Pc$ pairs are not as suppressed as they are in W+c and one may expect that contributions due to gluon splittings to $\Pc\bar\Pc$ pairs in Z+c can be much larger than in W+c. 

Understanding the behaviour of the SS selection is important since experimentally, the W+c-jet process is typically extracted by measuring an OS--SS cross section. 
The idea behind this extraction is that gluon splittings into charm-anticharm pairs diminish the sensitivity of the measurement to the strange-quark content of the proton, and are subtracted in the OS--SS selection. However, as we have seen throughout this work, additional effects like higher order corrections, off-diagonal CKM elements, etc.\ can be as large as the SS itself and tend to dilute this simple LO picture. A high-precision measurement of this process will therefore benefit from taking into account all effects quantified in the present work. 

In conclusion, in the present study we have shown that essentially all theoretical aspects of W+c-jet production at the LHC are under good theoretical control. The largest remaining sensitivity is to the PDFs which in turn may allow a precise extraction of the  strange-quark content of the proton based on new LHC data. To this end, all predictions obtained in this work are made publicly available\footnote{\url{https://www.precision.hep.phy.cam.ac.uk/results/hf-jets/}.
The predictions are available in the form of differential distributions.
At this url, additional predictions with different phase spaces at NNLO QCD accuracy are also available.}.

\section*{Acknowledgements}

We thank Zahari Kassabov for providing us with the reduced NNLO PDF sets.
The work of M.C.\ was supported by the Deutsche Forschungsgemeinschaft under grant 396021762 -- TRR 257.
The research of A.M., M.P., and R.P.\ has received funding from the European Research Council (ERC) under the European Union's Horizon 2020 Research and Innovation Programme (grant agreement no.~683211). A.M.\ was also supported by the UK STFC grants ST/L002760/1 and ST/K004883/1.
M.P.\ acknowledges support by the German Research Foundation (DFG) through the Research Training Group RTG2044 and through grant no INST 39/963-1 FUGG (bwForCluster NEMO) as well as the state of Baden-Württemberg through bwHPC.
R.P.\ acknowledges the support from the Leverhulme Trust and the Isaac Newton Trust.

\bibliographystyle{utphys.bst}
\bibliography{wc_nnlo}
\end{document}